\documentclass[10pt,letterpaper]{article}
\usepackage[top=0.85in,left=2.75in,footskip=0.75in]{geometry}

\usepackage{changepage}

\usepackage[utf8]{inputenc}

\usepackage{textcomp,marvosym}

\usepackage{fixltx2e}

\usepackage{graphicx}
\usepackage{caption}
\usepackage{subcaption}
\usepackage{amsmath,amssymb}
\usepackage[table,xcdraw]{xcolor}
\usepackage{multirow}
\usepackage{rotating}
\usepackage{cite}

\usepackage{nameref,hyperref}

\usepackage[right]{lineno}

\usepackage{microtype}
\DisableLigatures[f]{encoding = *, family = * }

\usepackage{rotating}
\usepackage{tabularx}
\newcommand{\fracc}[2]{\, \displaystyle \frac{ #1}{ #2}}

\newcommand{\all}[2]{\,\begin{align}
                   #1 
                    \label{#2}
                   \end{align}
}

\raggedright
\usepackage[aboveskip=1pt,labelfont=bf,labelsep=period,justification=raggedright,singlelinecheck=off]{caption}

\makeatletter
\renewcommand{\@biblabel}[1]{\quad#1.}
\makeatother

\date{}

\usepackage{lastpage,fancyhdr,graphicx}
\usepackage{epstopdf}
\fancyheadoffset[L]{2.25in}
\fancyfootoffset[L]{2.25in}
\begin{document}
\vspace*{0.35in}

\begin{flushleft}
{\Large
\textbf\newline{Qualities and Inequalities in Online Social Networks  through\\ the Lens of the Generalized Friendship Paradox }
}
\newline
\\
Naghmeh Momeni\textsuperscript{1,* },
Michael Rabbat\textsuperscript{1 },
\\
\bigskip
\bf{1} Department of Electrical and Computer Engineering, McGill University, 3480 University Street, Montr\'eal, Qu\'ebec, Canada, H3A 0E9
\\
\bigskip

%
%





* naghmeh.momenitaramsari@mail.mcgill.ca

\end{flushleft}
\section*{Abstract}
 
The friendship paradox is the phenomenon that in social networks, people on average have fewer friends than their friends do. The generalized friendship paradox is an extension to attributes other than the number of friends. The friendship paradox and its generalized version have gathered recent attention due to the information they provide about network structure and local inequalities. In this paper, we propose several measures of nodal qualities which capture different aspects of their activities and influence in online social networks. Using these measures we analyse the prevalence of the generalized friendship paradox over Twitter and we report high levels of prevalence (up to over 90\% of nodes). We contend that this prevalence of the friendship paradox and its generalized version arise because of the hierarchical nature of the connections in the network. This hierarchy is nested as opposed to being star-like. We conclude that these paradoxes are collective phenomena not created merely by a minority of well-connected or high-attribute nodes. Moreover, our results show that a large fraction of individuals can experience the generalized friendship paradox even in the absence of a significant correlation between degrees and attributes.


\section*{Introduction}

The \emph{friendship paradox} (FP), first introduced by Feld~\cite{feld1991}, is a phenomenon stemming  from the structural properties  of social  networks. It indicates that although most people think that they are more popular than their friends, in actuality the converse is true: on average, each person has fewer friends than his/her friends do. This observation sheds light on the local inequalities of social networks, how people organize their social ties, and how these inequalities extend to macro structures of social networks.  This paradox has also been observed in online settings as well~\cite{ugander2011,hodas2013}. It has been contended that this paradox can be exploited for early detection of flu outbreaks~\cite{christakis2010,garcia2014}, and more generally, finding well-connected nodes in  large networks~\cite{eom2014,han2012,lattanzi}.

The generalized version of the friendship paradox is an extension to attributes other than number of social ties. It  was introduced in~\cite{eom2014}, where it is shown that in scientific collaboration networks, each scholar has on average fewer citations than his/her collaborators do. The \emph{generalized friendship paradox} (GFP) has been studied analytically in \cite{jotunable,fotouhi2014}.  The GFP links intra-personal attributes to inter-personal ties, and thus sheds light on the interplay between nodal characteristics and network structure. It also takes a notable step towards characterizing the local inequalities of networks regarding non-structural nodal properties. In this paper, we study the GFP in the context of online social networks, and we consider nodal attributes that corresponds to \emph{influence}. 
We study how the structure of connections between nodes is related to the influence they have upon others. 

Finding highly-influential nodes in social networks is  useful for a variety of tasks such as understanding diffusion of information~\cite{kimura2007,kim2015social,kimura2010} and misinformation~\cite{budak2011,nguyen2012}, promoting cooperation~\cite{droz2009,szolnoki2009}, optimal product placement for marketing purposes~\cite{hill2006,kempe2003}, optimal immunization and vaccination strategies~\cite{perisic2009}, and studying the diffusion of innovation~\cite{song2007}. 
 
Characterizing influence is not straightforward. 
Being highly connected does not necessarily mean being influential. 
For example, in online social media, it has been found that highly-connected individuals are overwhelmed by information flows and sometimes cannot detect viral content effectively~\cite{hodas2012}. 
Thus, purely-structural measures alone (such as degree) cannot capture nodal influence. Furthermore, there are different types of influence, and nodes with different patterns of activities and impacts can be deemed influential. Most of the  online social networks include initiation and adoption processes. Each user can generate contents visible to other users who can re-post them. We need measures of influence that enable us to compare, for example, highly-active nodes with average clout with occasionally-active nodes with great clout.  

This paper proposes measures that capture multiple aspects of node activity and influence in online social networks.    Two of these measures quantify nodal activity, and four of them quantify inter-nodal influence. We compute and analyse  these measures on Twitter (the micro-blogging platform), but the measures are general and as we contend, they can be applied to any network setting with initiation and adoption mechanisms. We study the distributions and statistical properties of these measures.

In this paper, we  study the GFP (on the \emph{individual level}, in the terminology of~\cite{eom2014,jotunable}) through the lens of the measures of activity and influence that we introduce. We also introduce new measures for quantifying the GFP. Our measures assess to what extent nodes of a network experience the GFP. Throughout the paper, we use the term `Neighbor Superiority'~\cite{thezbabak} to refer to this phenomenon, because we found no evidence in the literature that, for example, most scholars think that they are more cited than their collaborators (which would contradict reality and create a `paradox'). Hence, the word paradox is not appropriate for contexts other than friendship. Furthermore, in online social media, evidence point towards the opposite direction~\cite{chou2012}, and most users assess their neighbors more highly than themselves.

We find high prevalence of neighbor superiority both in terms of connectivity and quality. For each of these nodal attributes, a vast majority of the nodes, even those who rank very highly in the population (for example, among the top 0.5\% in terms of tweeting activity, or in terms of popularity), experience neighbor superiority. We analyse the distributions of the measures and their prevalences more closely and uncover a hierarchical nature in the connectivity of the Twitter graph. We contend that neighbor superiority, and its special case, the friendship paradox,  are not mere mathematical artifacts that result from a star-like structure---a simplistic picture in which almost all nodes are connected to a few hubs, and these hubs make them experience neighbor superiority~\cite{feld1991}.  Instead, we show that there is a hierarchical nature in the pattern of connections in the network. Moreover, similar to the friendship paradox that enables biased sampling and detection of popular nodes~\cite{eom2014,garcia2014}, our results indicate that the same scheme can be applied in terms of non-structural nodal attributes  (i.e., qualities), to detect high-quality nodes.

The rest of this paper is organized as follows. We first introduce the terminology used throughout the paper. After describing the data set, we introduce measures of nodal quality (activity and influence), as well as measures to quantify neighbor superiority. We then present the distribution of nodal qualities and connectivity. We discuss results on neighbor superiority and focus on their implications for the underlying structure of the Twitter graph.

\section*{Terminology}
On Twitter, users can post short texts that should not exceed 140 characters. These posts are called \emph{tweets}. User can post \emph{original tweets}, or can repost another user's tweet, which is called \emph{retweeting}. Each user can \emph{follow} other users. When user A follows user B, we say that A is a \emph{follower} of B and B is a \emph{followee} of A. 
When A follows B, A subscribes to the tweets posted by B. Each user can see the tweets of his/her followees on his/her home Twitter feed.

 The underlying web of connectivities between users can be modelled as a graph. Users are mapped onto \emph{nodes}, and their connections onto \emph{links}. 
Note that on Twitter, B can follow A back or not. Mathematically, this means that the Twitter graph is \emph{directed}. 
There are two types of adjacency relationships that can be defined on the Twitter graph---follower and followee. We use the term  \emph{neighbor} to refer to both of these types of connection. So for each user, an neighbor can be either a follower or a followee. The number of followers and the number of followees of a user are called the \emph{in-degree} and \emph{out-degree} of that user, respectively. 

When user A posts a tweet, the followers of A can see it. When one of its followers, say B, retweets it, the followers of B can also see the tweet. With each retweet, the number of users who are exposed to the tweet increases. This is called a \emph{cascade} of the original tweet which was posted by A. The total number of retweets that a tweet by A receives is called the \emph{cascade size} of that tweet. Note that this retweet can be done either by A's own followers, or the followers of A's followers, and so on. After user B retweets one of A's tweets, then if a user C, which is a follower of B retweets that tweet, this retweet is counted only for the original tweet, and only for A. In other words, any cascade only has one root tweet and one initiating user. All the retweets are counted for that tweet and that user. This mechanism is internal to Twitter and we follow the same convention in this paper.

There are two categories of attributes that we consider in this paper. The first category consists of the in-degree and out-degree, which are \emph{structural} attributes. The second category  assesses tweeting activities of a node. We denote the attributes that belong to the latter category by \emph{quality} (Note that by \emph{quality} we mean an intrinsic fitness value that drives the connection and following patterns. It does not signify any quality of the content of the tweets). We define 6 different qualities in the section~\nameref{sec:measures}.

\section*{Data}
We use two datasets in this paper. The first one is presented by Yang and Leskovec~\cite{yang2011} and  contains over 470 million tweets by over 18 million users, which capture over 20\%  of all tweets posted over a 7-month period, starting from June 2009. For the network of connectivity of Twitter users, we use the dataset  collected  by Kwak et al.~\cite{kwak}. This dataset comprises all the  links between   users who
joined Twitter prior to  August 2009. We only consider users that are present in both data sets. The subgraph of connectivity has 5.8 million users and over 193 million links. The subset of all tweets that is considered includes over 200 million tweets.

\section*{Methods}
We discard repeated tweets for each user and count only the number of distinct tweets. Following the convention mentioned above, we count retweets only for the root user of a cascade, not for other users who retweet the message (who consequently received further retweets for their retweet). So the retweets for each user can be from those who directly follow the user, or the followers of the followers of the user, and so on.

%
%

\subsection*{Node Attributes as  Quality} \label{sec:measures}
Minding the specificities of the data set at hand, we considered six possible candidates as measures for node qualities. Two of these measures quantify the  activity of nodes, and four of them quantify their influence   on others.  Combining these six measures with in-degree and out-degree, we construct an 8-dimensional feature vector that characterizes each node. The six quality features are the following: 
\begin{enumerate}
\item   The \emph{number of tweets} (NT) is the total number of posts of the user, which includes original tweets and retweets. 
\item  The \emph{number of original tweets} (NOT) is the number of tweets that the user has initiated. 
\item  The \emph{total times retweeted} (TTR)  is the number of times that the posts initiated by the user got retweeted by other users.  It is the gross number of retweets that the user has received. 
 \item   The \emph{number of tweets retweeted} (NTR) is the number of tweets initiated by the user that received at least one retweet from other users. In other words, the NTR is the number of times that the user has created a cascade.  
 \item  The \emph{retweets per  tweet} (RPT) is the average number of retweets received by a tweet initiated by the user. In other words, the RPT of a user is the expected  cascade size that the user engenders. 
 \item  The \emph{fraction of tweets retweeted} (FTR) is the normalized version of NTR, that is, it is equal to the fraction of tweets initiated by the user that received at least one retweet from other users. This characterizes the clout of the user by assessing the likelihood that a tweet initiated by the user will engender a cascade.  
\end{enumerate}

As mentioned above, there are two categories: activity and influence. 
NT and NOT are measures of activity: NT  is a measure of \emph{total activity}; it measures how much a user posts tweets (that can be created by him/her or his/her peers). NOT is a measure of \emph{total novel activity}. 

The next four measures (TTR, NTR, RPT, and FTR) are measures of influence: TTR measures the \emph{total influence} of  a node  over followers.  NTR is a measure of \emph{total success}; it measures the number of successful initiations. RPT is a measure of \emph{efficiency}; it measures the expected influence per initiation. FTR is a measure of \emph{consistency}, which measures the likelihood of generating cascade of any size per initiation. 

Note that these measures are general, and need not be confined to Twitter. NT and NOT can be used for any social network in which a specific action can be defined as `activity' (e.g., posting content on Google$+$, Pinterst, Instagram, Tumblr, etc.). TTR, NTR, RPT and FTR can be used in any network context in which there is a mechanism for sharing, reposting, or adoption. For example, on Pinterest, users can \emph{pin} items on their boards, and their followers can \emph{re-pin} them (equivalent of retweeting). On Facebook, posts can be \emph{shared} and on Tumblr, users can \emph{re-blog} the posts by other users.

Through a hypothetical example, we  shed light on the nuances of these measures and the different aspects of  the users they capture. Let us consider three users: user 1 has made 100 original tweets, one of them has received 1000 retweets and the rest have received none. User 2 has posted 100 original tweets; each of them have received 10 retweets. So in total, user 2 has received 1000 retweets. User 3 has made only 10 tweets;, each of them have received 50 retweets. So user 3 has received 500 retweets in total.

  Users 1 and 2 have equal TTR values of 1000, which exceeds that of user 3 (which is 500).   Note that TTR cannot distinguish between the first two users, while their patterns of influence are clearly different. We can distinguish between user 1 and 2 using FTR, because  the FTR of user 1 is 0.01, whereas the FTR of user 2  is equal to 1.  In this example   user 1 has had a moment in the sun, and does not have a steady influence over other users, whereas user 2 consistently creates cascades (of smaller size as compared to that of user 1). In other words, user 2 is more reliable to engender cascades than user 1, but the cascade is not as large. 
  
 Now let us consider users 2 and 3. For both of them, the FTR is 1, which means that for both users, every tweet has has been retweeted at least once. However, the RPT of user 3 is 50, and the RPT of user 2 is 10. This means that although user 3 is not as active as user 2, the cascades created by user 3 are on average five times larger.

We now turn our attention to the distribution of these different measures of quality in the network. In addition to the inequalities of the degrees, we investigate the inequalities between node qualities that exist in the network.

\subsection*{Measures of neighbor Superiority}
In this section we introduce measures to quantify neighbor superiority, which is the essence of the friendship paradox and its generalized version. These measures are generalizations of the measures we introduced in~\cite{momeni2014}, which pertain to undirected networks only. 

In  the present paper, the network under consideration is directed. However, to develop intuition about the measures that we are going to introduce, first let us consider a simple undirected network. So there is no follower/followee distinction; rather, each node simply has neighbors. In this case, we can compare the degree of each node with,  say, the average of the degrees of its neighbors. If the degree of the node is smaller than the average of its neighbors' degree, we say that the node is experiencing mean  neighbor superiority. Throughout the network, different nodes with different degrees can be experiencing average neighbor superiority.  A question we can ask about the network under study is that, how large should the degree of a node be so that it will not experience mean neighbor superiority? To address this question, we introduce the notion of \emph{critical degree for the mean}, which is defined to be  the maximum of the degrees of all the nodes in the network that experience mean neighbor superiority. In other words, no node in the network with degree greater than the critical degree experiences average neighbor superiority. Let us denote the set of neighbors of node $x$ by $N_x$, and let us denote the degree of node $x$ by $k_x$. The critical degree for the mean can be expressed as follows:
\all{
\widetilde{K} = 
\max \bigg\{k_x \bigg| k_x < \fracc{\sum_{y\in N_x}k_y}{|N_x|}    \bigg\}
.}{Kdef}

A drawback of using the mean neighbor degree is that a node might be experiencing neighbor superiority only because one of its many neighbors had a very large degree, hence making the mean neighbor degree large. We can also compare the degree of each node with the median neighbor degree, which alleviates the problem of outliers (see~\cite{momeni2014,kooti2014}).  A node is said to be experiencing median neighbor superiority if more than half of its neighbors have higher degrees than it does.  Let us denote the median by $M(\cdot)$. Similar to the mean, we can also define the \emph{critical degree for the median} as follows: 
\all{
\widehat{K} = 
\max \bigg\{k_x \bigg| k_x < M \left(k_y | y\in N_x \right)      \bigg\}
.}{Kdef}

Note that any nodal attribute can be compared to the mean or median of the neighbors in order to define neighbor superiority. It need not be degree. It can be age, for example. 

In addition to critical values, another way of quantifying neighbor superiority in the network would be to measure the prevalence of neighbor superiority, that is the fraction of nodes who experience a given type of neighbor superiority. For example, we can ask what fraction of nodes experience median neighbor superiority or mean neighbor superiority. 

Now let us extend these definitions to the case of a directed network. In this case, each node has two distinct sets of neighbors: followers and followees. Thus, we can compare each attribute of a node to its followers and its followees. The results of these comparisons need not be the same (in fact,  as we will demonstrate, they are not). This proliferates the number of ways we can compare a node to its neighbors. For example, we can compare the in-degree  of a node (i.e., number of followers) with the  in-degree of its followers, or the in-degree of its followees. We can also use measures of quality, as introduced above. For the same reasoning as mentioned for the undirected case, we can use both the mean and the median for comparison. This engenders several possible ways of defining neighbor superiority, as well as corresponding critical values (both for the mean and median versions). 


In total, there are 32 different critical values that can be defined:  median/mean   follower/followee   superiority for 8 different possible attributes.  
Corresponding to these 32 different types of superiorities, we can also measure 32 fractions that reflect what fraction of nodes in the network experience a given type of superiority.

\section*{Results and Discussion}

\subsection*{Distribution of Quality and Degree}
We computed the 6 measures of quality, as well as in-degree and out-degree, for all the nodes in the Twitter network. The distribution of all the 8 nodal attributes are highly skewed. Table~\ref{tab:summary} presents the summary statistics of the distributions.  The percentage headers indicate percentiles: e.g., the first row of the 90\% column is 41, which means that 90\% of the users have in-degree less than or equal to 41.   It is of note that the four zeros at the bottom of the third column indicate that  over 75\% of the users  never got retweeted, which implies  a high skew in the distribution. 
In other words,  on Twitter, most people only  observe. They read, but seldom retweet. 

 \begin{table}[!h]
\begin{tabular}{cccccccc}
 & Mean  & Median (50\%)& 75\% & 90\%  & 95\%  & 99\%  & Max    (100\%)                      \\ \hline\hline
 
In-Degree                               & 35.2  & 4      & 13   & 41    & 92    & 459   & 625520                          \\
\rowcolor[HTML]{CBCEFB}
Out-Degree                              & 35.2  & 10     & 22   & 57    & 115   & 539   & 86800                           \\

NT                        & 37.3  & 5      & 22   & 79    & 157   & 526   & 85316                           \\
\rowcolor[HTML]{CBCEFB} 
NOT               & 35.1  & 4      & 20   & 74    & 148   & 499   & 85234                           \\

TTR                        & 2.40  & 0      & 0    & 2     & 6     & 35    & 82036                           \\
\rowcolor[HTML]{CBCEFB} 
NTR             & 0.83  & 0      & 0    & 1     & 3     & 13    & 4803                             \\
RPT             & 0.167 & 0      & 0    & 0.056 & 0.167 & 1.224 & 12567.0\\
\rowcolor[HTML]{CBCEFB} 
FTR& 0.02  & 0      & 0    & 0.036 & 0.100 & 0.500 & 1.0                            
\\ \hline
\end{tabular} \caption{Summary statistics of nodal attributes. }
\label{tab:summary}
\end{table}

The distribution of the 8 nodal attributes is depicted in Fig~1. For each attribute, we divide the interval between the minimum and maximum values of the attribute into 100 bins. The bin sizes increase logarithmically and the distributions are plotted on a log-log scale. Note that for each bin the logarithm of the endpoint is evaluated. All nodal attributes exhibit a heavy-tailed distribution. Figs~1g and~1h exhibit interesting behavior: the majority of the nodes (over 80\%) have zero RPT and zero FTR. Setting these users aside, the rest of the population exhibit distributions for RPT and FTR that, unlike other 6 nodal attributes,  are not monotonically decreasing.

\begin{figure}[!h]
        \centering
        \begin{subfigure}[b]{0.5 \columnwidth}
                \includegraphics[width=\columnwidth, height=50mm]{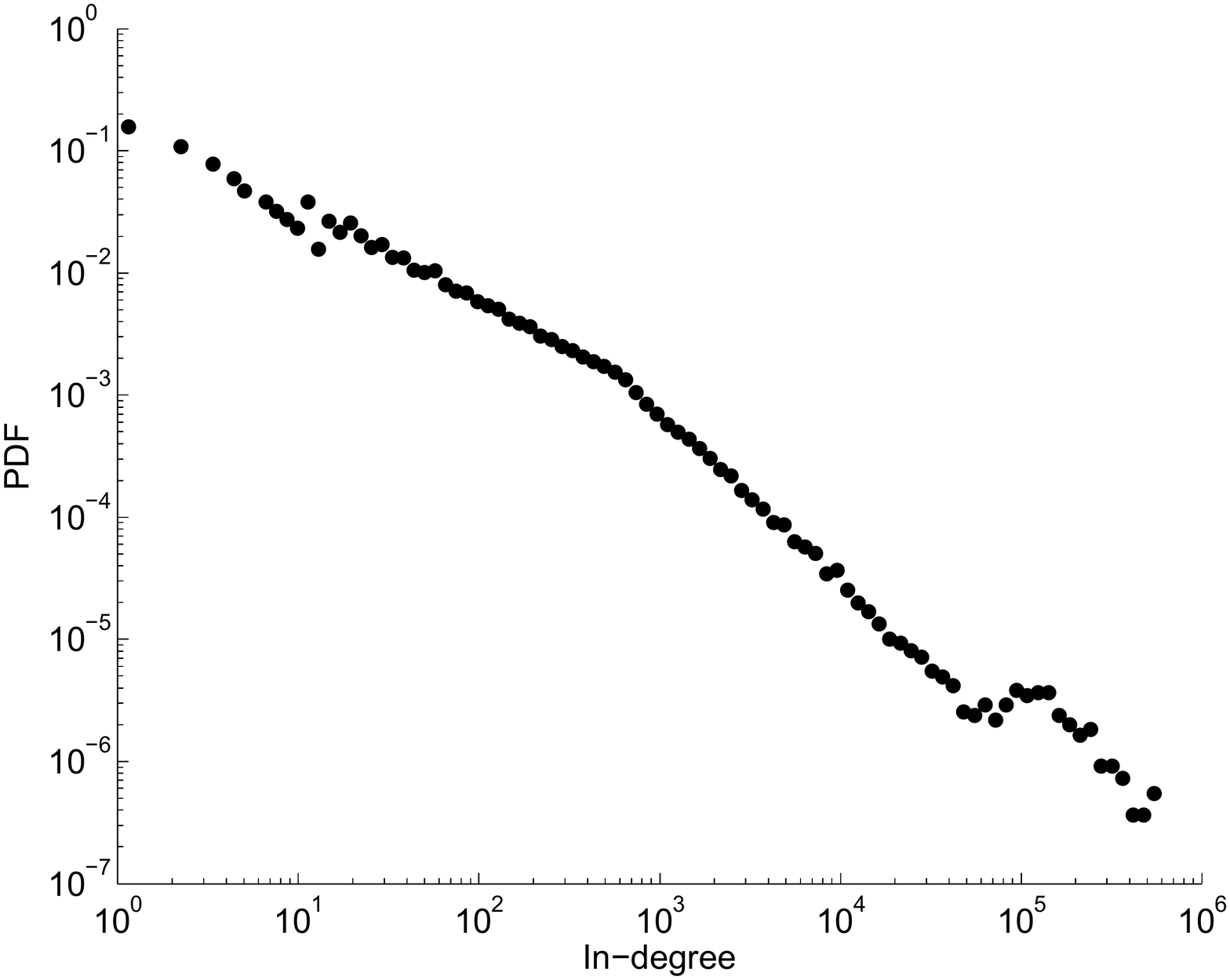}
                \caption{In-degree}
                \label{InDegreeD}
        \end{subfigure}%
        ~ 
        \begin{subfigure}[b]{0.5 \columnwidth}
                \includegraphics[width=\columnwidth, height=50mm]{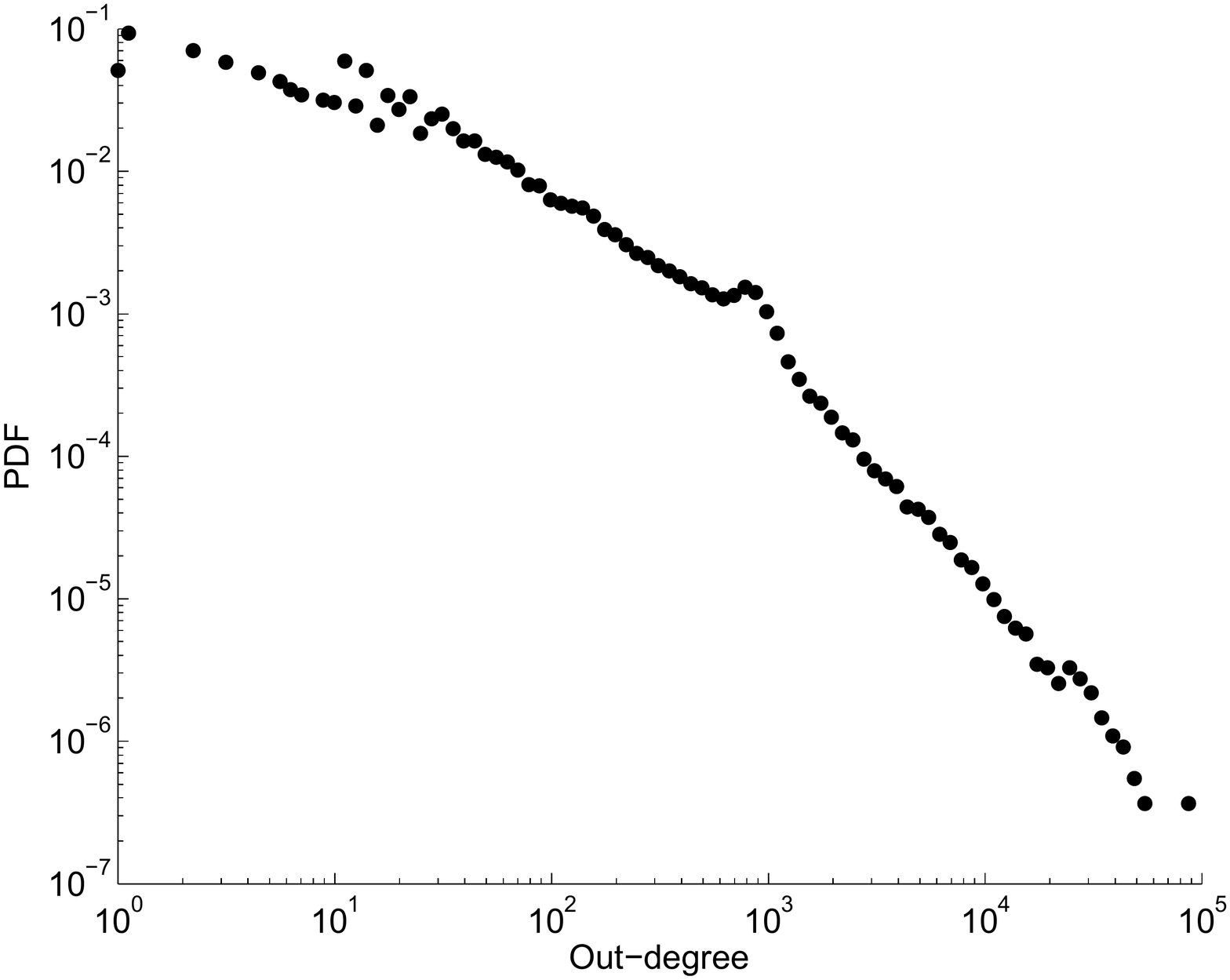}
                \caption{Out-degree }
                \label{OutdegreeD}
        \end{subfigure}
        \\
              \centering
        \begin{subfigure}[b]{0.5\textwidth}
                \includegraphics[width=\textwidth]{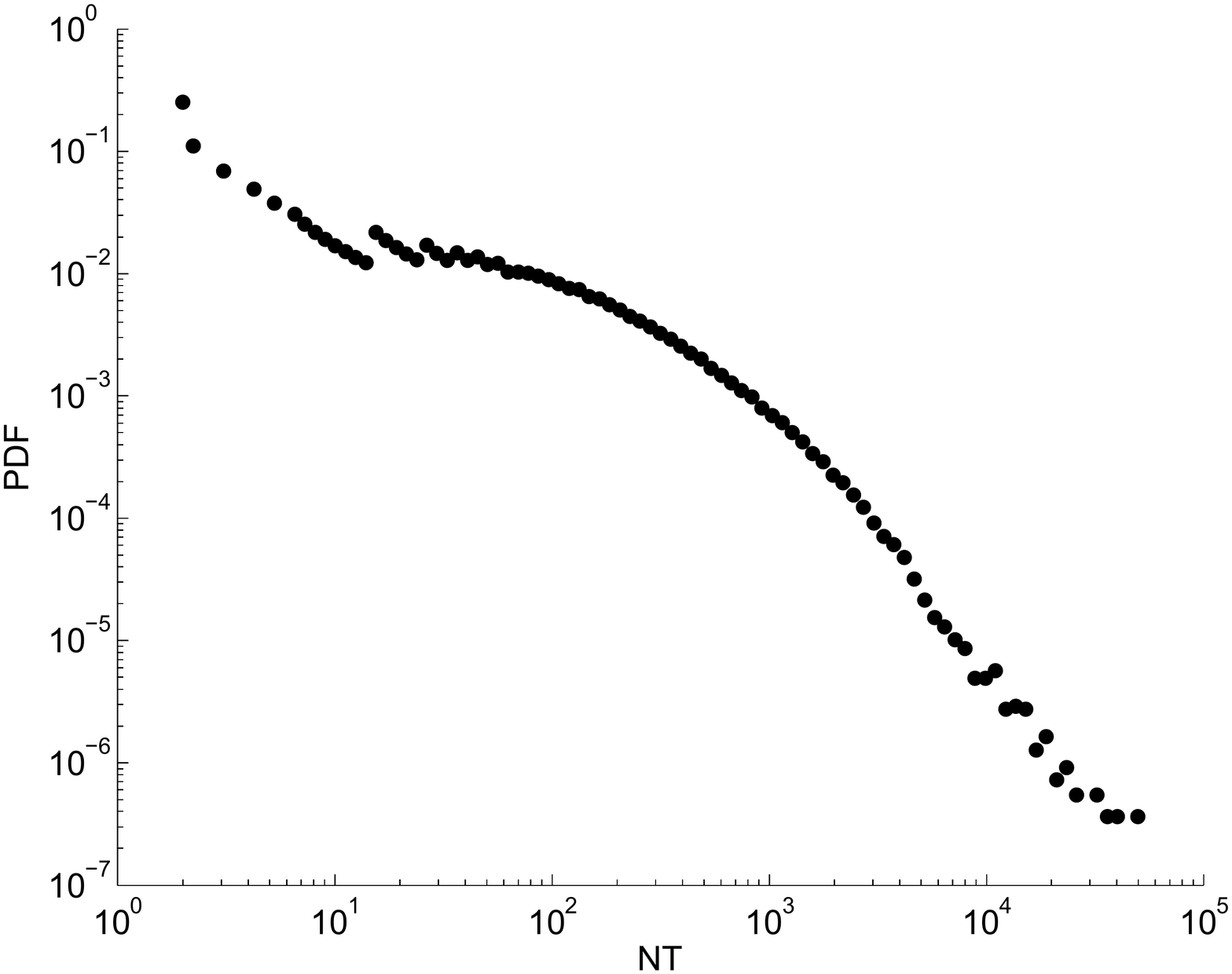}
                \caption{NT: Number of Tweets}
                \label{NTD}
        \end{subfigure}%
        ~ 
        \begin{subfigure}[b]{0.5\textwidth}
                \includegraphics[width=\textwidth]{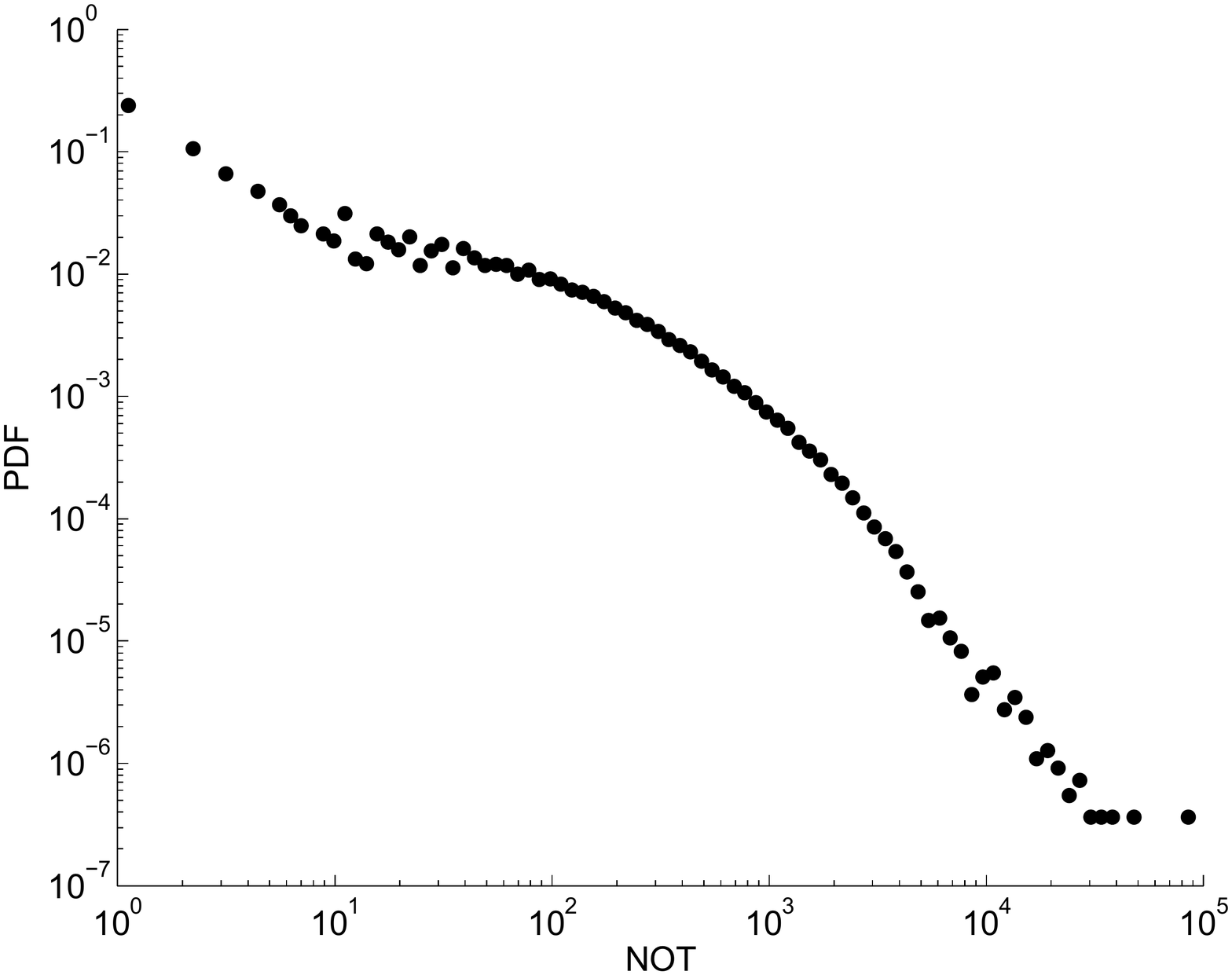}
                \caption{NOT: Number of Original Tweets.}
                \label{NOTD}
        \end{subfigure}
        \\
         \centering
        \begin{subfigure}[b]{0.5\textwidth}
                \includegraphics[width=\textwidth, height=50mm]{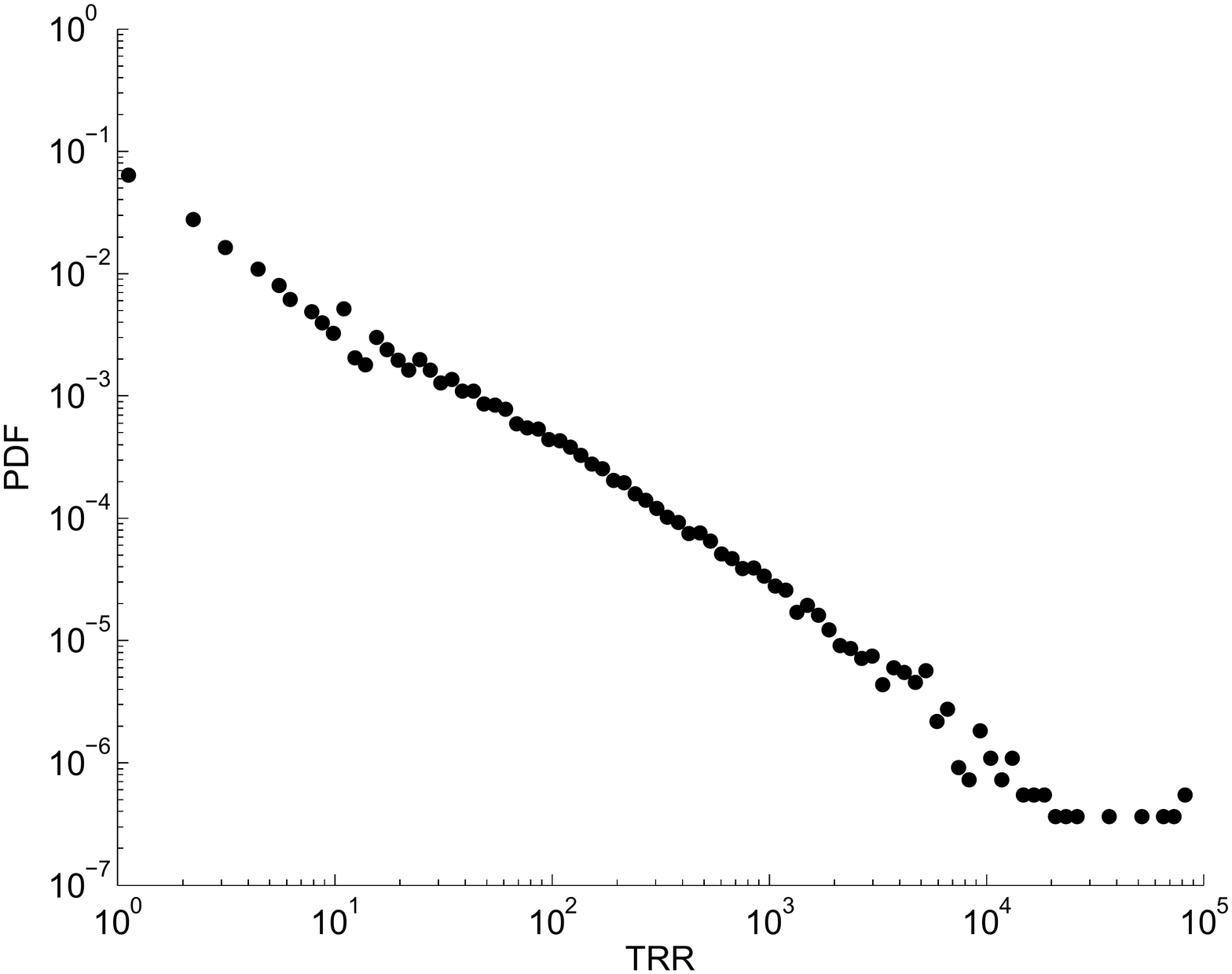}
                \caption{TTR: Total Times Retweeted.}
                \label{TTR}
        \end{subfigure}%
        ~ 
        \begin{subfigure}[b]{0.5\textwidth}
                \includegraphics[width=\textwidth, height=50mm]{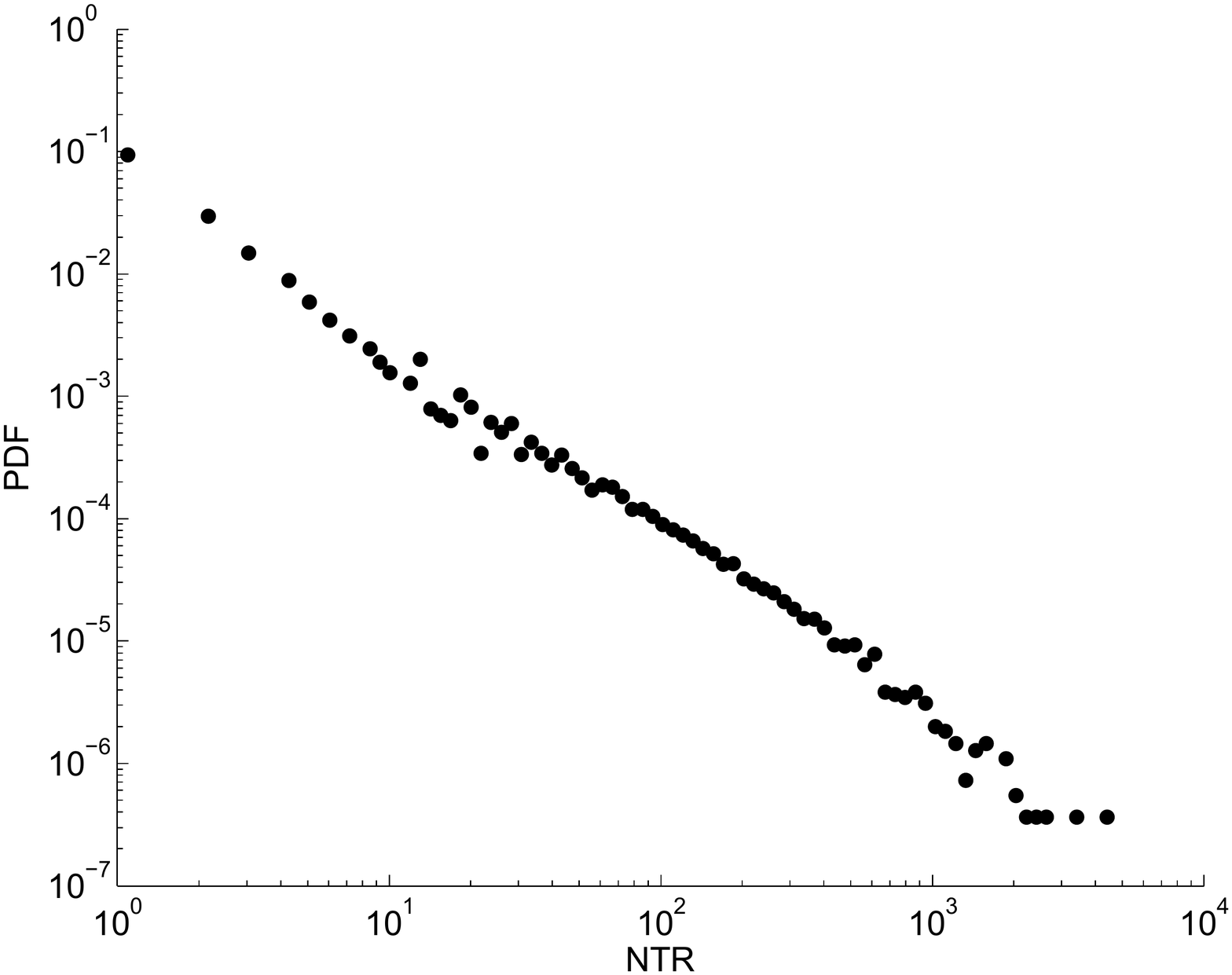}
                \caption{NTR: Number of Tweets Retweeted. }
                \label{NTRD}
        \end{subfigure}
        \\
         \centering
        \begin{subfigure}[b]{0.5\textwidth}
                \includegraphics[width=\textwidth]{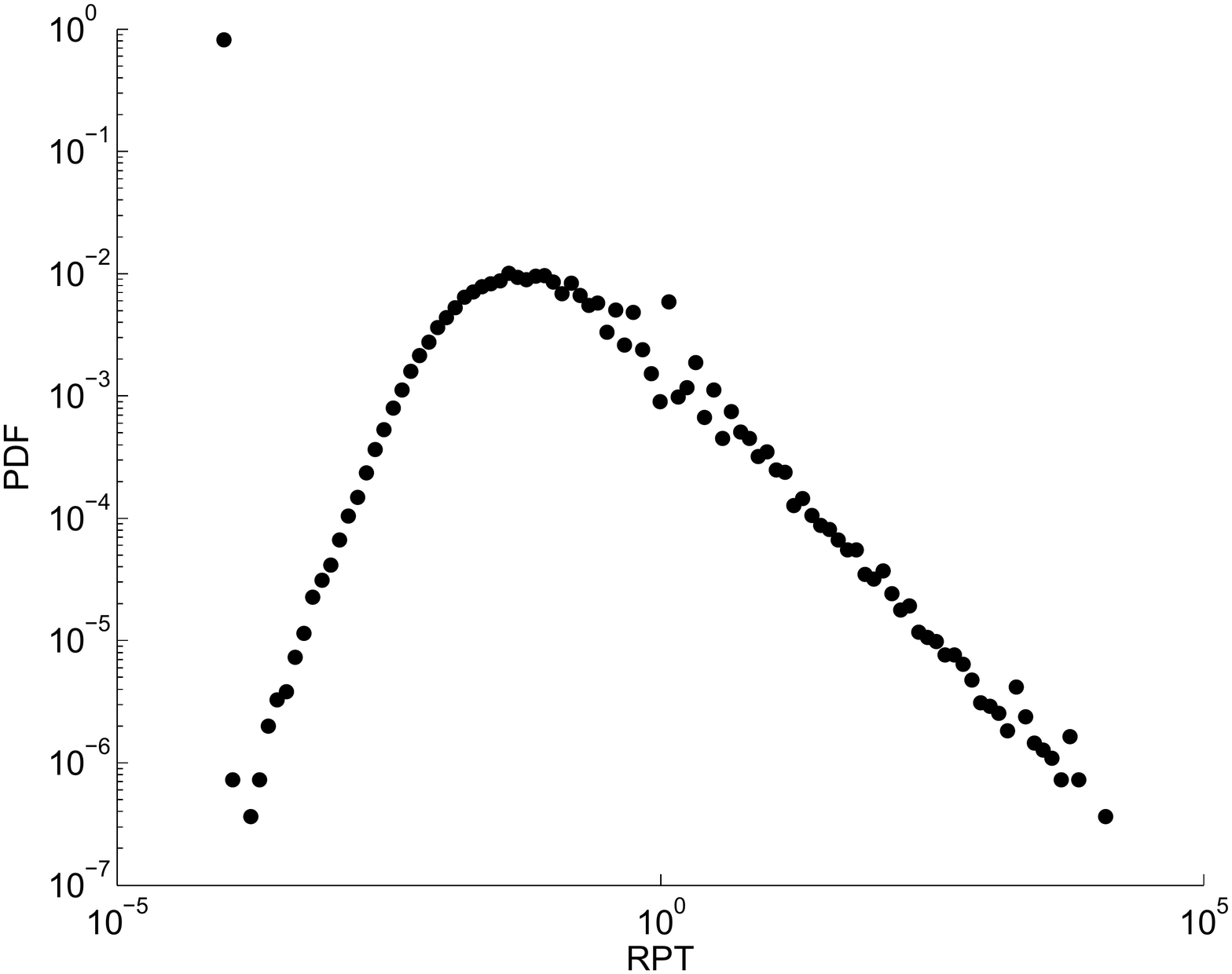}
                \caption{RPT: Retweets Per Tweet.}
                \label{RPTD}
        \end{subfigure}%
        \begin{subfigure}[b]{0.5\textwidth}
                \includegraphics[width=\textwidth]{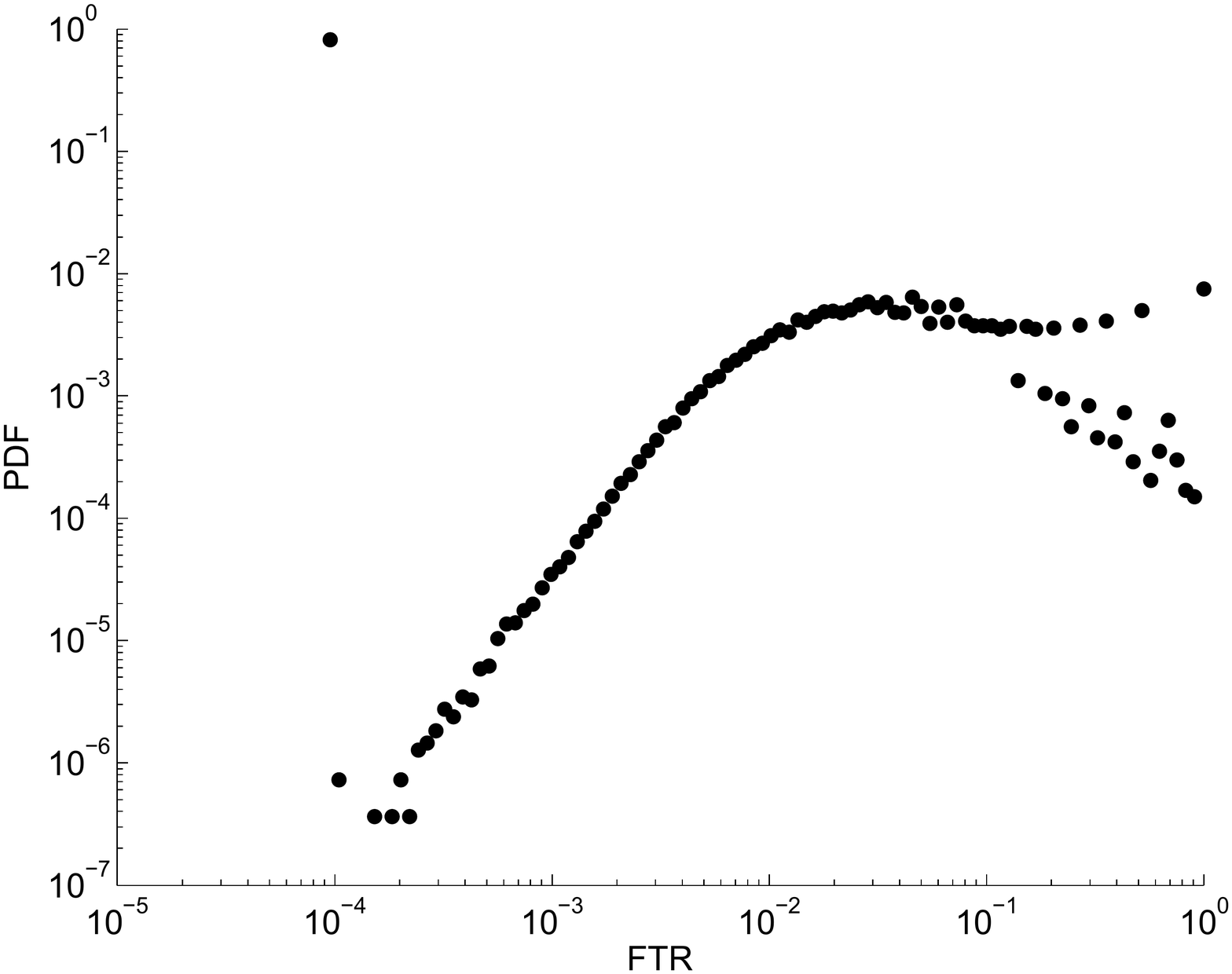}
                \caption{FTR: Fraction of Tweets Retweeted. }
                \label{FTRD}
        \end{subfigure}
        \caption{\footnotesize{Distributions of different nodal attributes.} }\label{fig:hist}
\end{figure}

\begin{table}[]
\centering
\begin{tabular}{ccccccccc}
                         & In                                             & Out                                            & NT                                             & NOT                                            & TTR                                            & NTR                                            & \multicolumn{1}{l}{RPT}                        & \multicolumn{1}{l}{FTR}                        \\ \cline{2-9} 
\multicolumn{1}{c|}{In}  & \multicolumn{1}{c|}{\cellcolor[HTML]{CBCEFB}1} & \multicolumn{1}{c|}{0.271}                     & \multicolumn{1}{c|}{0.041}                     & \multicolumn{1}{c|}{0.040}                     & \multicolumn{1}{c|}{0.067}                     & \multicolumn{1}{c|}{0.112}                     & \multicolumn{1}{c|}{0.002}                     & \multicolumn{1}{c|}{0.002}                     \\ \cline{2-9} 
\multicolumn{1}{c|}{Out} & \multicolumn{1}{c|}{0.271}                     & \multicolumn{1}{c|}{\cellcolor[HTML]{CBCEFB}1} & \multicolumn{1}{c|}{0.128}                     & \multicolumn{1}{c|}{0.121}                     & \multicolumn{1}{c|}{0.042}                     & \multicolumn{1}{c|}{0.156}                     & \multicolumn{1}{c|}{0.002}                     & \multicolumn{1}{c|}{0.053}                     \\ \cline{2-9} 
\multicolumn{1}{c|}{NT}  & \multicolumn{1}{c|}{0.041}                     & \multicolumn{1}{c|}{0.128}                     & \multicolumn{1}{c|}{\cellcolor[HTML]{CBCEFB}1} & \multicolumn{1}{c|}{0.993}                     & \multicolumn{1}{c|}{0.108}                     & \multicolumn{1}{c|}{0.486}                     & \multicolumn{1}{c|}{-0.001}                    & \multicolumn{1}{c|}{0.013}                     \\ \cline{2-9} 
\multicolumn{1}{c|}{NOT} & \multicolumn{1}{c|}{0.040}                     & \multicolumn{1}{c|}{0.121}                     & \multicolumn{1}{c|}{0.993}                     & \multicolumn{1}{c|}{\cellcolor[HTML]{CBCEFB}1} & \multicolumn{1}{c|}{0.086}                     & \multicolumn{1}{c|}{0.419}                     & \multicolumn{1}{c|}{-0.002}                    & \multicolumn{1}{c|}{-0.002}                    \\ \cline{2-9} 
\multicolumn{1}{c|}{TTR} & \multicolumn{1}{c|}{0.067}                     & \multicolumn{1}{c|}{0.042}                     & \multicolumn{1}{c|}{0.108}                     & \multicolumn{1}{c|}{0.086}                     & \multicolumn{1}{c|}{\cellcolor[HTML]{CBCEFB}1} & \multicolumn{1}{c|}{0.231}                     & \multicolumn{1}{c|}{0.356}                     & \multicolumn{1}{c|}{0.054}                     \\ \cline{2-9} 
\multicolumn{1}{c|}{NTR} & \multicolumn{1}{c|}{0.112}                     & \multicolumn{1}{c|}{0.156}                     & \multicolumn{1}{c|}{0.486}                     & \multicolumn{1}{c|}{0.419}                     & \multicolumn{1}{c|}{0.231}                     & \multicolumn{1}{c|}{\cellcolor[HTML]{CBCEFB}1} & \multicolumn{1}{c|}{0.003}                     & \multicolumn{1}{c|}{0.112}                     \\ \cline{2-9} 
\multicolumn{1}{c|}{RPT} & \multicolumn{1}{c|}{0.002}                     & \multicolumn{1}{c|}{0.002}                     & \multicolumn{1}{c|}{-0.001}                    & \multicolumn{1}{c|}{-0.002}                    & \multicolumn{1}{c|}{0.356}                     & \multicolumn{1}{c|}{0.003}                     & \multicolumn{1}{c|}{\cellcolor[HTML]{CBCEFB}1} & \multicolumn{1}{c|}{0.075}                     \\ \cline{2-9} 
\multicolumn{1}{c|}{FTR} & \multicolumn{1}{c|}{0.002}                     & \multicolumn{1}{c|}{0.053}                     & \multicolumn{1}{c|}{0.013}                     & \multicolumn{1}{c|}{0.002}                     & \multicolumn{1}{c|}{0.054}                     & \multicolumn{1}{c|}{0.112}                     & \multicolumn{1}{c|}{0.075}                     & \multicolumn{1}{c|}{\cellcolor[HTML]{CBCEFB}1} \\ \cline{2-9} 
\end{tabular}
\caption{Correlation coefficients between nodal attributes}
\label{correlations}
\end{table}

Table~\ref{correlations} presents the correlation coefficients between the six measures of influence, as well as in-degree and out-degree (28 pairs in total). It can be observed that the the magnitude of correlation between  21 pairs are below 0.15, and only one is above 0.5. This confirms that these measures capture distinct components of nodal attributes. Also note that in-degree and out-degree are not highly correlated with measures of influence. This is a notable observation, and further confirms that measures of connectivity are not necessarily correlated with influence. Had the correlation coefficient between degrees and nodal attributes been large, then the GFP would become an artifact of the FP. However, this is not the case, and they cannot be ascribed to a common cause. Finally, note that NT and NOT are highly correlated,  which is  expected. We have included both  because in the calculation of other quantities NOT was employed.

\subsection*{Results on  neighbor Superiority}
The fraction of nodes in the network that experience the corresponding types of neighbor superiority  are presented in Table~\ref{tab:frac}. The critical values pertaining to each type of neighbor superiority is presented in Table~\ref{tab:crit}. In both tables, the first two rows pertain to the structural attributes and the other rows pertain to the nodal attributes considered as quality.

\begin{table}[!h]
\centering
\resizebox{.95\textwidth}{!}{%
\begin{tabular}{cccccccccc}
 &  &  & \multicolumn{7}{c}{Critical Values} \\ \cline{4-10} 
 &  &  & \multicolumn{3}{c}{Mean} &  & \multicolumn{3}{c}{Median} \\ \cline{4-10} 
 &  &  & Follower &  & Followee &  & Follower &  & Followee \\ \hline\hline
\multicolumn{2}{c}{} & \cellcolor[HTML]{FFFFFF}In-degree & \cellcolor[HTML]{FFFFFF}2890 & \cellcolor[HTML]{FFFFFF} & \cellcolor[HTML]{FFFFFF}155657 & \cellcolor[HTML]{FFFFFF} & \cellcolor[HTML]{FFFFFF}1894 & \cellcolor[HTML]{FFFFFF} & \cellcolor[HTML]{FFFFFF}114629 \\
\multicolumn{2}{c}{\multirow{-2}{*}{Structural}} & \cellcolor[HTML]{CBCEFB}Out-degree & \cellcolor[HTML]{CBCEFB}2887 & \cellcolor[HTML]{CBCEFB} & \cellcolor[HTML]{CBCEFB}2887 & \cellcolor[HTML]{CBCEFB} & \cellcolor[HTML]{CBCEFB}2108 & \cellcolor[HTML]{CBCEFB} & \cellcolor[HTML]{CBCEFB}1997 \\ \hline
\multicolumn{2}{c}{} & \cellcolor[HTML]{FFFFFF}NT & \cellcolor[HTML]{FFFFFF}3305 & \cellcolor[HTML]{FFFFFF} & \cellcolor[HTML]{FFFFFF}5009 & \cellcolor[HTML]{FFFFFF} & \cellcolor[HTML]{FFFFFF}1837 & \cellcolor[HTML]{FFFFFF} & \cellcolor[HTML]{FFFFFF}5009 \\
\multicolumn{2}{c}{\multirow{-2}{*}{\begin{tabular}[c]{@{}c@{}}Quality: \\ Activity\end{tabular}}} & \cellcolor[HTML]{CBCEFB}NOT & \cellcolor[HTML]{CBCEFB}2853 & \cellcolor[HTML]{CBCEFB} & \cellcolor[HTML]{CBCEFB}5009 & \cellcolor[HTML]{CBCEFB} & \cellcolor[HTML]{CBCEFB}1827 & \cellcolor[HTML]{CBCEFB} & \cellcolor[HTML]{CBCEFB}5009 \\ \hline
\multicolumn{2}{c}{} & \cellcolor[HTML]{FFFFFF}TTR & \cellcolor[HTML]{FFFFFF}540 & \cellcolor[HTML]{FFFFFF} & \cellcolor[HTML]{FFFFFF}2590 & \cellcolor[HTML]{FFFFFF} & \cellcolor[HTML]{FFFFFF}628 & \cellcolor[HTML]{FFFFFF} & \cellcolor[HTML]{FFFFFF}1962 \\
\multicolumn{2}{c}{} & \cellcolor[HTML]{CBCEFB}NTR & \cellcolor[HTML]{CBCEFB}219 & \cellcolor[HTML]{CBCEFB} & \cellcolor[HTML]{CBCEFB}301 & \cellcolor[HTML]{CBCEFB} & \cellcolor[HTML]{CBCEFB}141 & \cellcolor[HTML]{CBCEFB} & \cellcolor[HTML]{CBCEFB}286 \\
\multicolumn{2}{c}{} & \cellcolor[HTML]{FFFFFF}RPT & \cellcolor[HTML]{FFFFFF}54.1 & \cellcolor[HTML]{FFFFFF} & \cellcolor[HTML]{FFFFFF}64.5 & \cellcolor[HTML]{FFFFFF} & \cellcolor[HTML]{FFFFFF}40.0 & \cellcolor[HTML]{FFFFFF} & \cellcolor[HTML]{FFFFFF}19.0 \\
\multicolumn{2}{c}{\multirow{-4}{*}{\begin{tabular}[c]{@{}c@{}}Quality: \\ Influence\end{tabular}}} & \cellcolor[HTML]{CBCEFB}FTR & \cellcolor[HTML]{CBCEFB}0.975 & \cellcolor[HTML]{CBCEFB} & \cellcolor[HTML]{CBCEFB}0.911 & \cellcolor[HTML]{CBCEFB} & \cellcolor[HTML]{CBCEFB}0.975 & \cellcolor[HTML]{CBCEFB} & \cellcolor[HTML]{CBCEFB}0.896 \\ \hline
\end{tabular}
}
\caption{Critical values for different types of neighbor superiority.}
\label{tab:crit}
\end{table}

A hierarchy of connections can be discerned from Table~\ref{tab:frac}. For any given type of neighbor superiority, either in the mean or the median version, we observe that the fraction of nodes experiencing followee superiority exceeds the fraction of nodes experiencing follower superiority (which is true in 15 out of all 16 possible cases). 
This suggests the existence of a hierarchy of attachment, which is a result of the tendency of users to follow those who have higher attributes than them, both in terms of degree (in/out) and quality.

\begin{table}[!h]
\centering
\resizebox{.95\textwidth}{!}{%
\begin{tabular}{cccccccccc}
 &  &  & \multicolumn{7}{c}{Fraction of nodes experiencing neighbor superiority (\%)} \\ \cline{4-10} 
 &  &  & \multicolumn{3}{c}{Mean} &  & \multicolumn{3}{c}{Median} \\ \cline{4-10}
 &  &  & Follower &  & Followee &  & Follower &  & Followee \\ \hline\hline
\multicolumn{2}{c}{} & \cellcolor[HTML]{FFFFFF}In-degree & \cellcolor[HTML]{FFFFFF}85.5 & \cellcolor[HTML]{FFFFFF} & \cellcolor[HTML]{FFFFFF}93.7 & \cellcolor[HTML]{FFFFFF} & \cellcolor[HTML]{FFFFFF}79.7 & \cellcolor[HTML]{FFFFFF} & \cellcolor[HTML]{FFFFFF}90.2 \\
\multicolumn{2}{c}{\multirow{-2}{*}{Structural}} & \cellcolor[HTML]{CBCEFB}Out-degree & \cellcolor[HTML]{CBCEFB}86.1 & \cellcolor[HTML]{CBCEFB} & \cellcolor[HTML]{CBCEFB}92.5 & \cellcolor[HTML]{CBCEFB} & \cellcolor[HTML]{CBCEFB}82.0 & \cellcolor[HTML]{CBCEFB} & \cellcolor[HTML]{CBCEFB}80.7 \\ \hline
\multicolumn{2}{c}{} & \cellcolor[HTML]{FFFFFF}NT & \cellcolor[HTML]{FFFFFF}71.4 & \cellcolor[HTML]{FFFFFF} & \cellcolor[HTML]{FFFFFF}87.2 & \cellcolor[HTML]{FFFFFF} & \cellcolor[HTML]{FFFFFF}58.4 & \cellcolor[HTML]{FFFFFF} & \cellcolor[HTML]{FFFFFF}79.3 \\
\multicolumn{2}{c}{\multirow{-2}{*}{\begin{tabular}[c]{@{}c@{}}Quality: \\ Activity\end{tabular}}} & \cellcolor[HTML]{CBCEFB}NOT & \cellcolor[HTML]{CBCEFB}71.2 & \cellcolor[HTML]{CBCEFB} & \cellcolor[HTML]{CBCEFB}87.2 & \cellcolor[HTML]{CBCEFB} & \cellcolor[HTML]{CBCEFB}57.8 & \cellcolor[HTML]{CBCEFB} & \cellcolor[HTML]{CBCEFB}79.4 \\ \hline
\multicolumn{2}{c}{} & \cellcolor[HTML]{FFFFFF}TTR & \cellcolor[HTML]{FFFFFF}65.9 & \cellcolor[HTML]{FFFFFF} & \cellcolor[HTML]{FFFFFF}83.3 & \cellcolor[HTML]{FFFFFF} & \cellcolor[HTML]{FFFFFF}33.0 & \cellcolor[HTML]{FFFFFF} & \cellcolor[HTML]{FFFFFF}67.8 \\
\multicolumn{2}{c}{} & \cellcolor[HTML]{CBCEFB}NTR & \cellcolor[HTML]{CBCEFB}65.2 & \cellcolor[HTML]{CBCEFB} & \cellcolor[HTML]{CBCEFB}83.1 & \cellcolor[HTML]{CBCEFB} & \cellcolor[HTML]{CBCEFB}32.5 & \cellcolor[HTML]{CBCEFB} & \cellcolor[HTML]{CBCEFB}67.2 \\
\multicolumn{2}{c}{} & \cellcolor[HTML]{FFFFFF}RPT & \cellcolor[HTML]{FFFFFF}64.4 & \cellcolor[HTML]{FFFFFF} & \cellcolor[HTML]{FFFFFF}81.9 & \cellcolor[HTML]{FFFFFF} & \cellcolor[HTML]{FFFFFF}34.2 & \cellcolor[HTML]{FFFFFF} & \cellcolor[HTML]{FFFFFF}66.9 \\
\multicolumn{2}{c}{\multirow{-4}{*}{\begin{tabular}[c]{@{}c@{}}Quality: \\ Influence\end{tabular}}} & \cellcolor[HTML]{CBCEFB}FTR & \cellcolor[HTML]{CBCEFB}63.0 & \cellcolor[HTML]{CBCEFB} & \cellcolor[HTML]{CBCEFB}80.4 & \cellcolor[HTML]{CBCEFB} & \cellcolor[HTML]{CBCEFB}34.0 & \cellcolor[HTML]{CBCEFB} & \cellcolor[HTML]{CBCEFB}66.5 \\ \hline
\end{tabular}
}
\caption{Fraction of nodes experiencing different types of neighbor superiority. }
\label{tab:frac}
\end{table}

For all types of mean and median followee superiority the fractions of nodes experiencing the corresponding neighbor superiority are above 80\% and 66\%, respectively. 
Table~\ref{tab:frac} shows that the fraction of nodes experiencing 12 out of 16 types of median superiority is higher than 57\%. This means that for these users, more than half of the users they are connected to have higher attributes than them. 
This challenges the simplistic picture that reduces neighbor superiority to a mere statistical artifact. This simplistic picture contends that neighbor superiority merely results from the existence of a few well-connected nodes with high attributes that make their neighbors experience mean neighbor superiority by lifting their neighbor-averaged attributes. We observe that for most of the nodes, it is not a single dominant neighbor that makes them experience neighbor superiority; rather, it is more than half of their neighbors that do this collectively.

 
The values of bottom half of the column pertaining to median follower superiority are smaller, as compared to other figures in Table~\ref{tab:frac}. Note that these four correspond to the four zeros in the 75\% percentile column of Table~\ref{tab:summary}. We can explain this observation saying that  for a large majority of the network (over 75\%), the values of TTR, NTR, RPT and FTR    are  equal to zero. It is plausible to deduce that these values are also zero for  the majority of the followers of each of these users. This renders the median value of their followers equal to zero, which makes them not experience follower superiority. However, since even one nonzero follower suffices to lift the mean above zero, the fraction of nodes experiencing mean follower superiority is much larger than those experiencing median follower superiority (the range of fractions for the mean version is between 63\% and 66\%, whereas for the median version, the range is between 32\% and 34\%). Comparing these fractions with the corresponding figures on the rightmost column of Table~\ref{tab:frac} provides further evidence for the existence of a following hierarchy. In short, users rarely follow down, they mostly tend to follow up or across.

The critical values also provide insight into the hierarchical structure of the connectivity of the Twitter network. For example, for  the mean followee  superiority in   in-degree, the critical value of 155657. This means that even a user who has 155657 followers follows users who on average have more followers than him/her. Noting that only 1\% of the users have more than 460 followers (Table~\ref{tab:summary}), the user with 155657 ranks very highly in terms of number of followers (99.99 percentile), and even this user is following those who on average have more followers. 
Similarly, for the median followee superiority in in-degree, the critical value is 114629. This means that even  for a user with this many followers, the majority of his/her followees have more followers than s/he does. 

In addition to degrees, similar observations can be made on nodal qualities. Consider TTR as an example. For the mean followee superiority in TTR, the critical value is 2590. Note that, as Table~\ref{tab:summary} presents, more than 75\% of all the users never get retweeted, that is, they have TTR of zero. Even a user with such a  high value of TTR experiences mean followee superiority. Similarly, for RPT, Table~\ref{tab:summary} tells us that 95\% of all the users have RPT values below 0.16. From Table~\ref{tab:crit} we observe that the critical values for the mean followee superiority in RPT is 64.5. Even a user with RPT as large as 64.5 follows users that are on average more influential in terms of cascade size. These provide evidence for the hierarchical nature of the connections of the Twitter network both in terms of connectivity and in terms of nodal qualities.

Fig~2 illustrates the empirical distribution of experiencing superiority pertaining to different attributes. The horizontal axis is log-scaled for better visibility. We construct 50 bins in each case, and for all the users who fall in each bin, we calculated the fraction of them who experience the given type of superiority.

\begin{figure}[!h]
        \centering
        \begin{subfigure}[b]{0.5 \columnwidth}
                \includegraphics[width=\columnwidth, height=50mm]{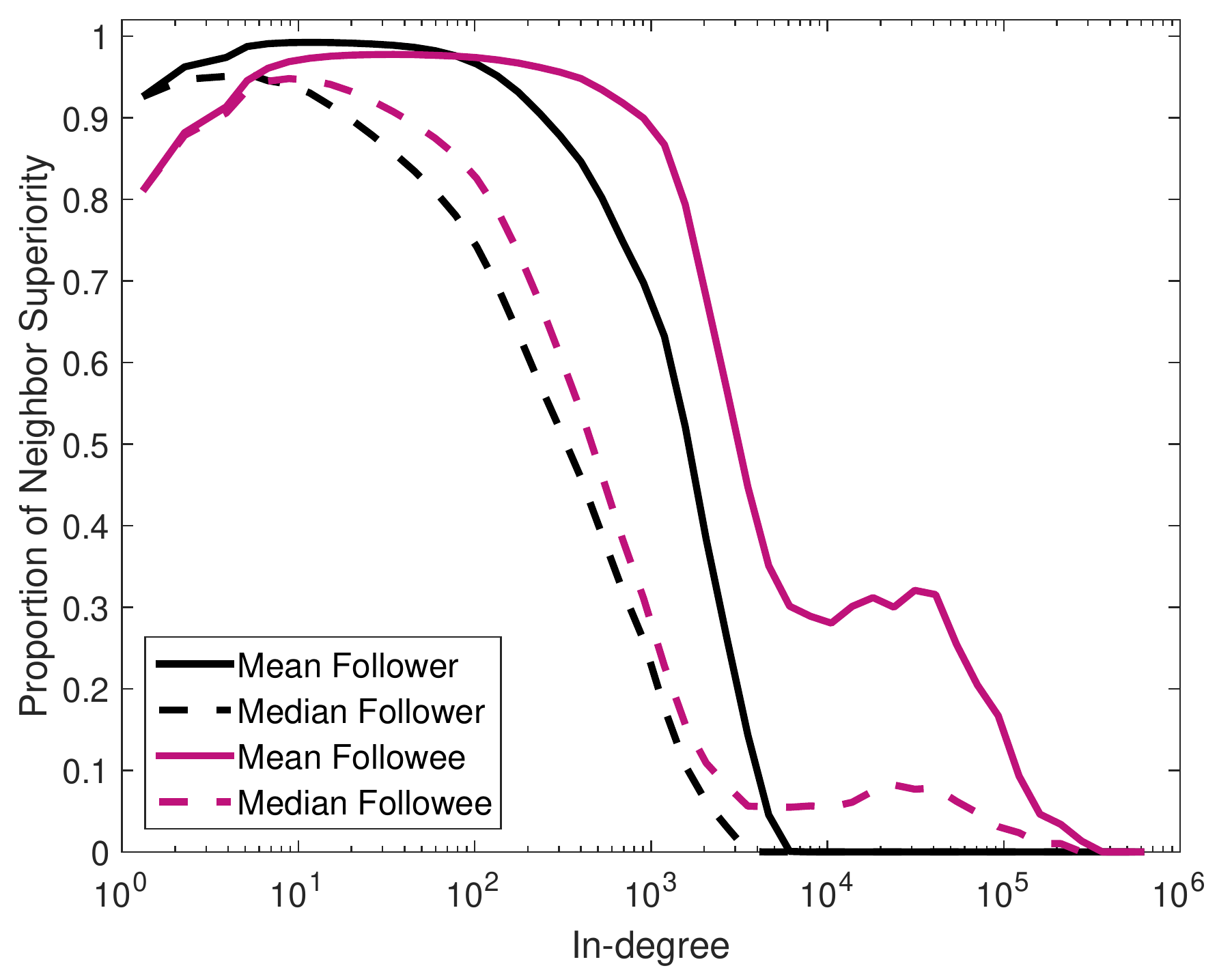}
                \caption{In-degree}
                \label{InDegree}
        \end{subfigure}%
        ~ 
        \begin{subfigure}[b]{0.5 \columnwidth}
                \includegraphics[width=\columnwidth, height=50mm]{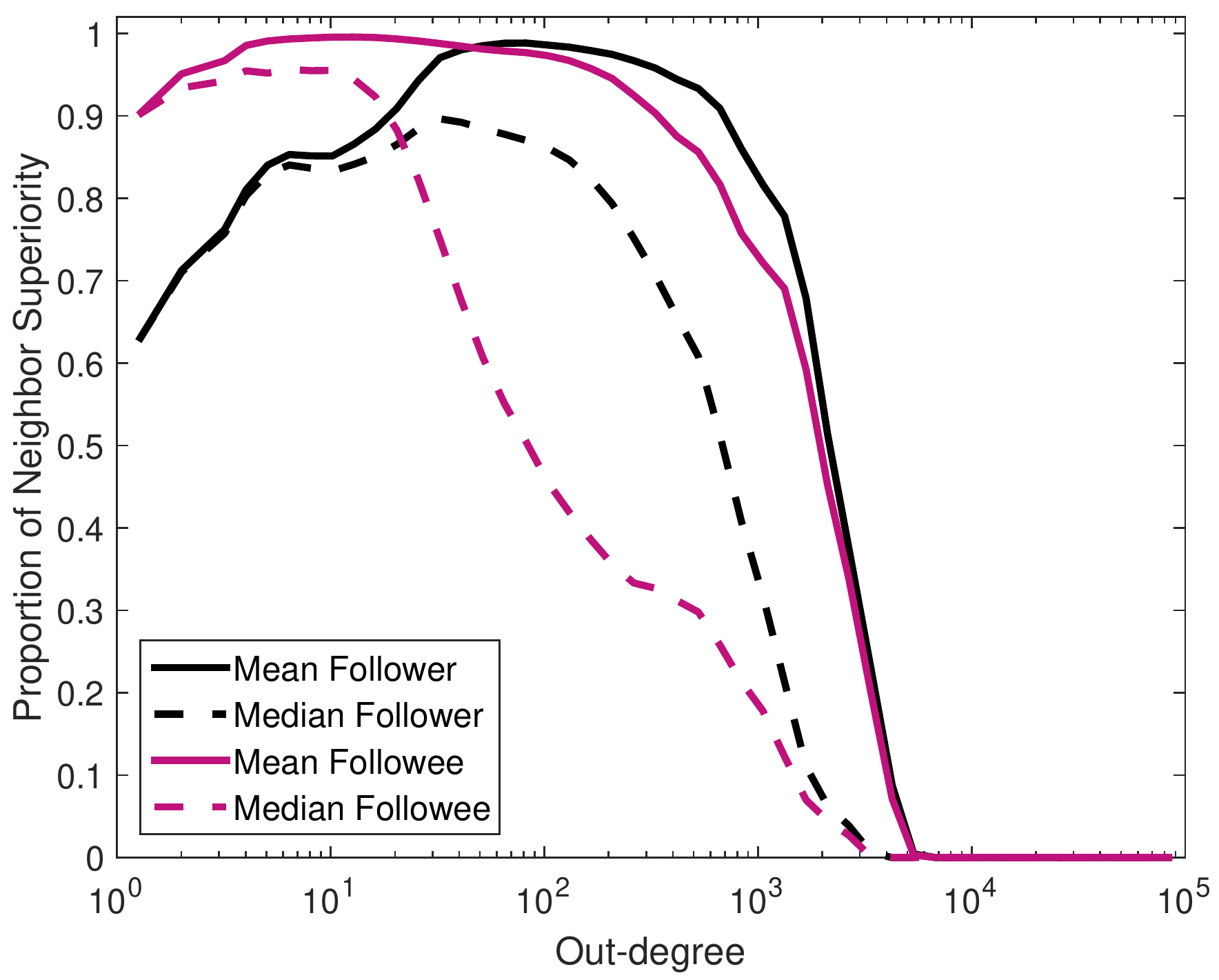}
                \caption{Out-degree}
                \label{Outdegree}
        \end{subfigure}
        \\
              \centering
        \begin{subfigure}[b]{0.5\textwidth}
                \includegraphics[width=\textwidth, height=50mm]{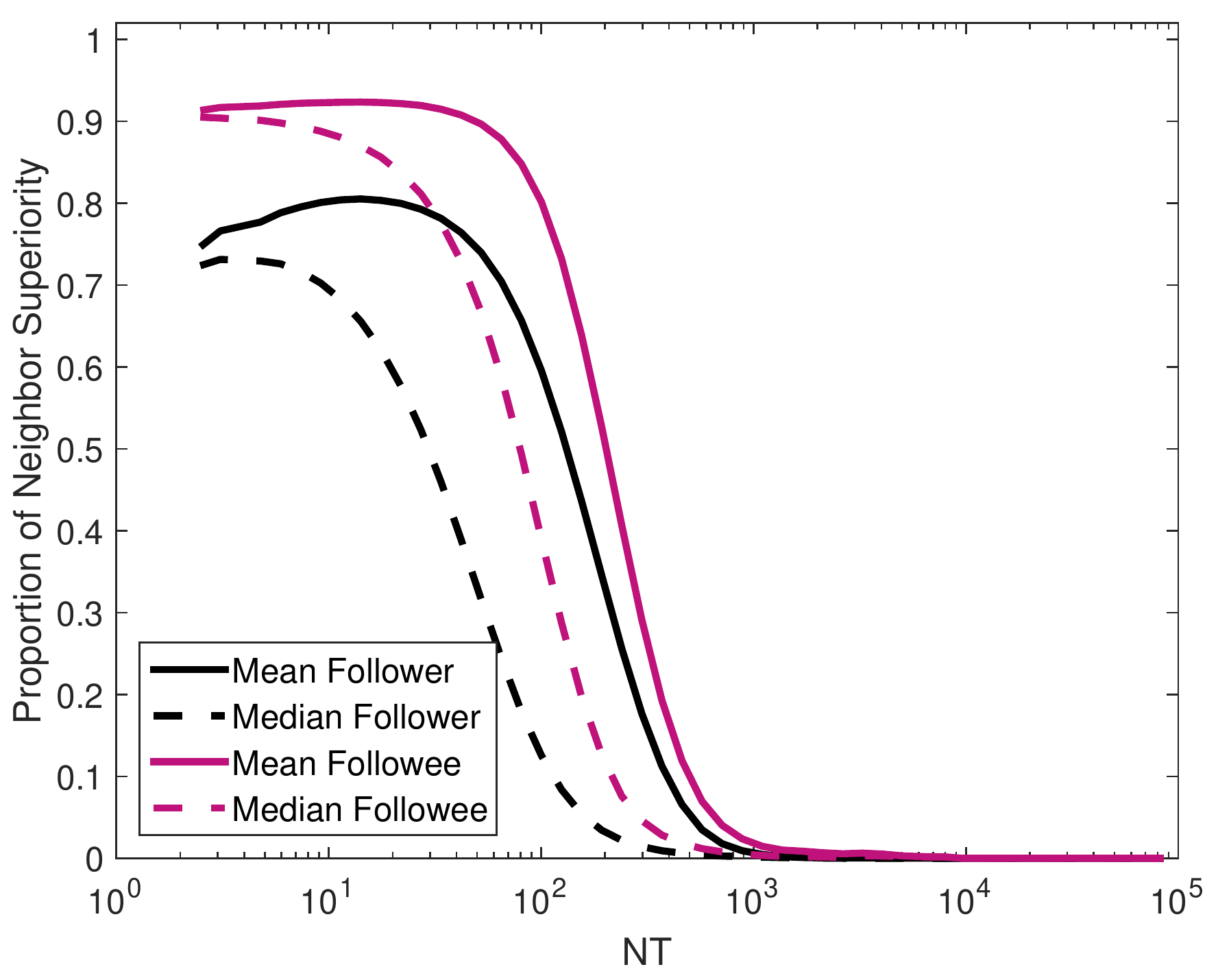}
                \caption{NT: Number of Tweets}
                \label{NT}
        \end{subfigure}%
        ~ 
        \begin{subfigure}[b]{0.5\textwidth}
                \includegraphics[width=\textwidth, height=50mm]{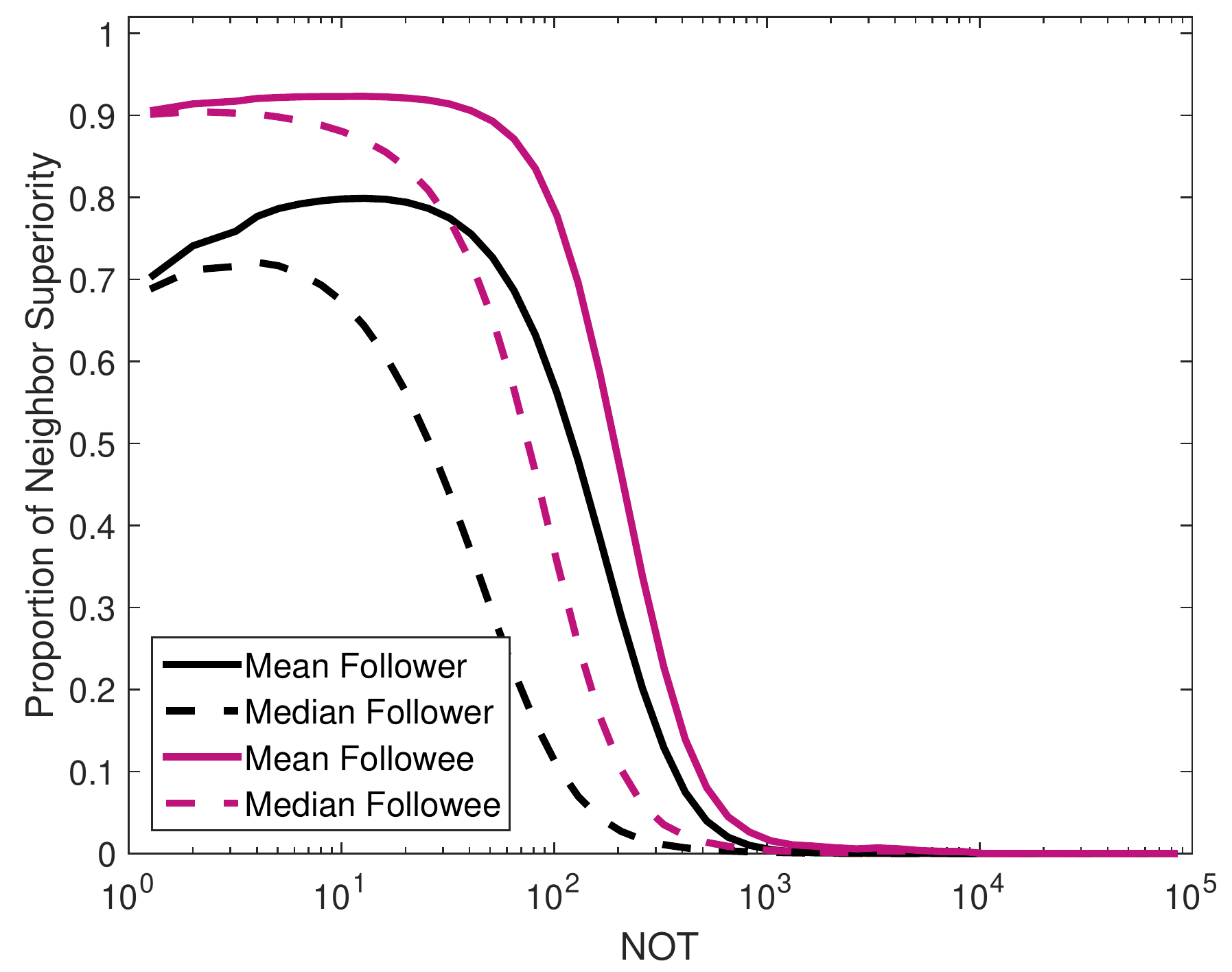}
                \caption{NOT: Number of Original Tweets.}
                \label{NOT}
        \end{subfigure}
        \\
         \centering
        \begin{subfigure}[b]{0.5\textwidth}
                \includegraphics[width=\textwidth, height=50mm]{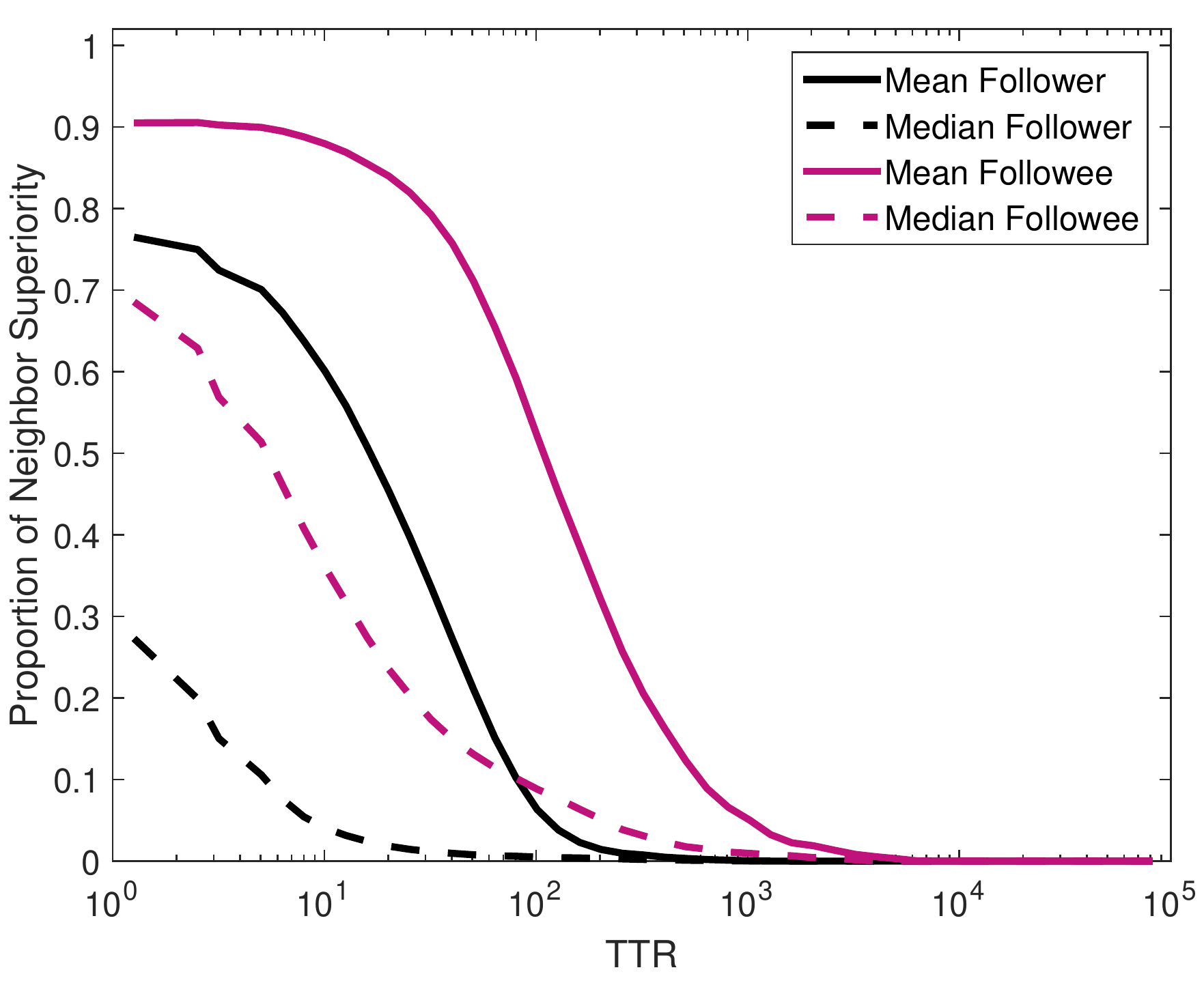}
                \caption{TTR: Total Times Retweeted.}
                \label{TTR}
        \end{subfigure}%
        ~ 
        \begin{subfigure}[b]{0.5\textwidth}
                \includegraphics[width=\textwidth, height=50mm]{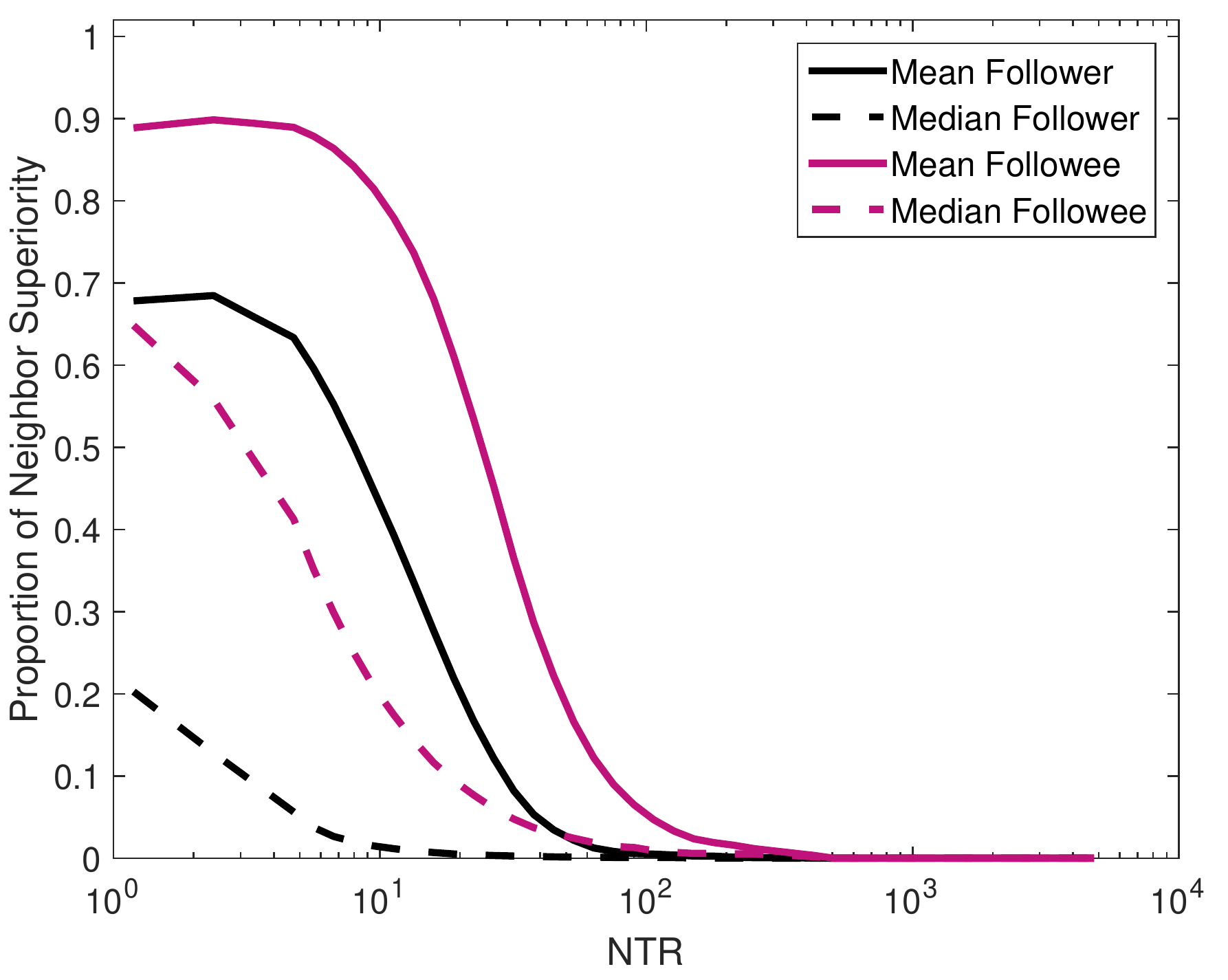}
                \caption{NTR: Number of Tweets Retweeted. }
                \label{NTR}
        \end{subfigure}
        \\
         \centering
        \begin{subfigure}[b]{0.5\textwidth}
                \includegraphics[width=\textwidth]{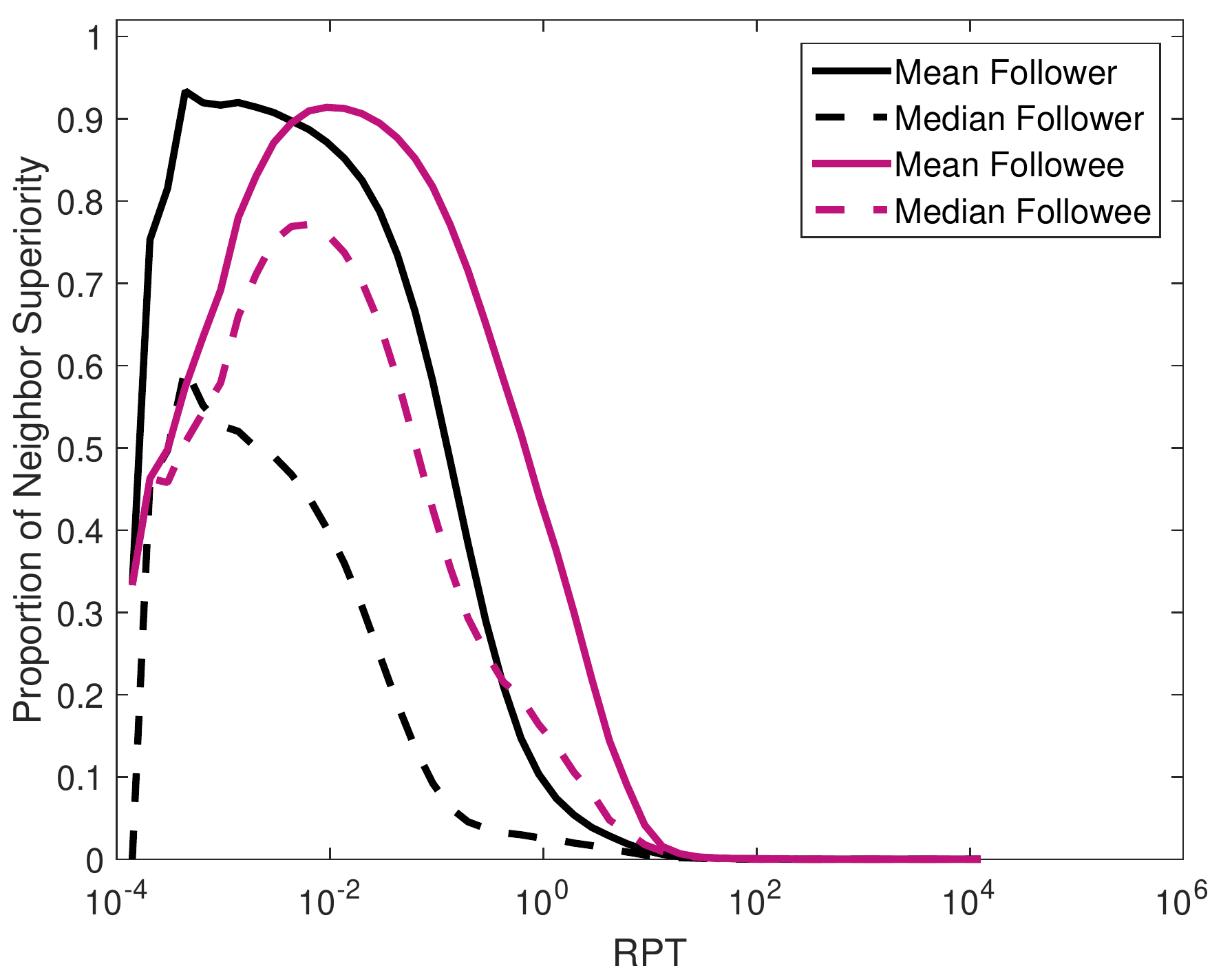}
                \caption{RPT: Retweets Per Tweet.}
                \label{RPT}
        \end{subfigure}%
        ~ 
        \begin{subfigure}[b]{0.5\textwidth}
                \includegraphics[width=\textwidth]{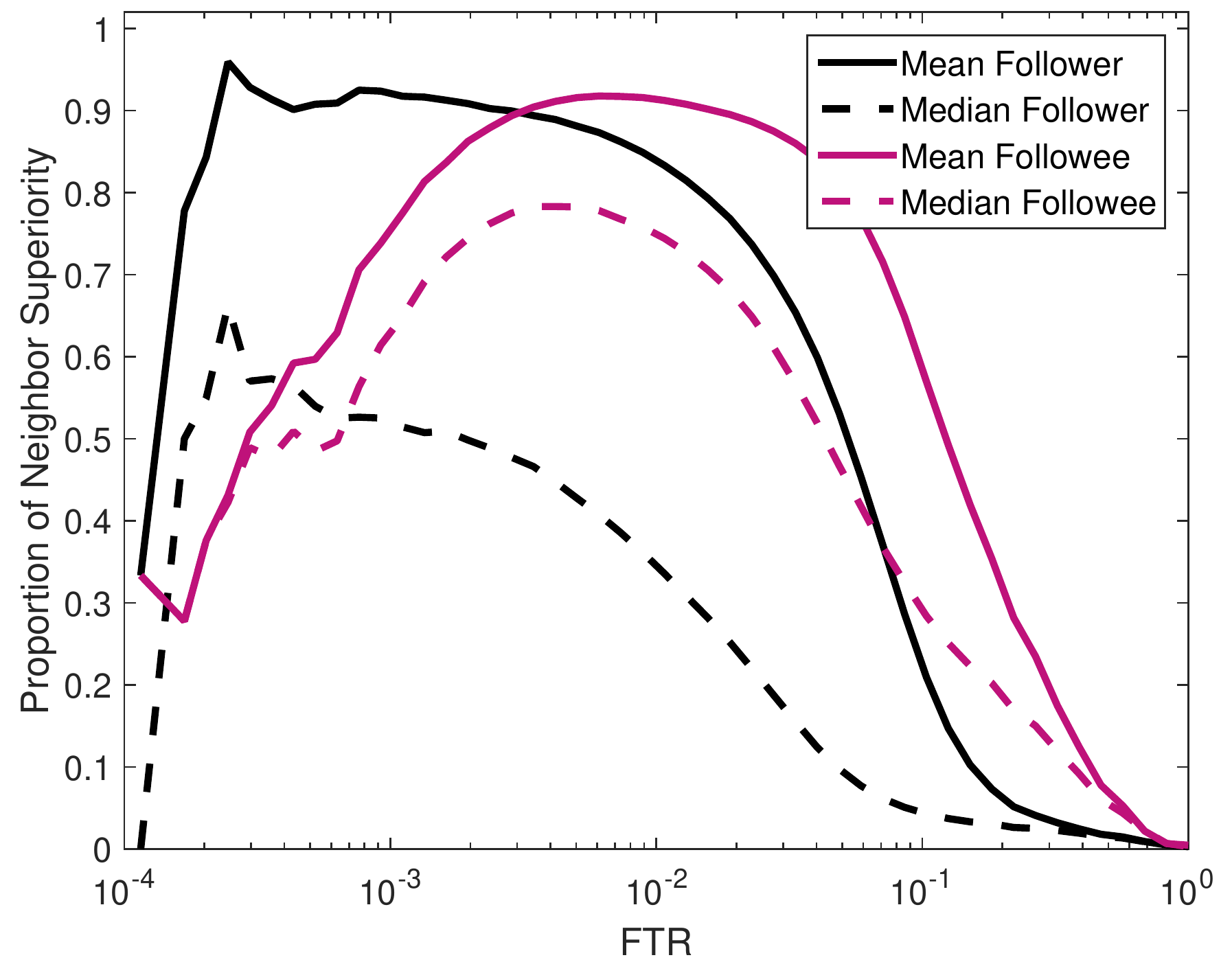}
                \caption{FTR: Fraction of Tweets Retweeted. }
                \label{FTR}
        \end{subfigure}
        \caption{\footnotesize{Proportion of  neighbor superiority of different types for different nodal attributes.} }\label{fig:prob}
\end{figure}

The intuitive expectation for any type of superiority might be that the higher the attribute of a node is, the less likely it would be for that node to experience neighbor superiority. In other words, one might expect to observe a uniformly decreasing likelihood of experiencing neighbor superiority as a function of any nodal attribute, which  is maximized when the value of the attribute takes its minimum values. However, in Fig~2, only two subfigures resemble this scheme, which are Figs~2e and~2f. All other subfigures present either curves with plateaus, or unimodal ones. This is a crucial observation with interesting consequences, as we shall discuss in detail next. Note that the point where the curves hit the x-axis are pertained to the corresponding critical values.

Let us consider Fig~2a as an example, and let us first focus on the solid red curve. For the minimum degree,   we observe that that proportion of experiencing mean followee superiority is 0.8, which means that 20\% of the users with minimum in-degree do not experience mean followee superiority. The proportion  \emph{increases} to above 0.95 when the degree is 5, then  follows a wide  plateau up to degree of 1000, and then decreases. The proportion of experiencing followee superiority is 0.3 for an in-degree as high as 20000: of the users with around 20000 followers, 30\% follow users with, on average, more followers. So critical values do not pertain to outliers. In other words,  a high critical value cannot be construed as ``maybe this is the only node among all nodes with high values of that attribute that experiences neighbor superiority". Rather, a high critical value suggests a continuous decline in the proportion of experiencing neighbor superiority. A high critical value does suggest that in the network under consideration, even the nodes  with high values of the attribute experience neighbor superiority. 

Note that Fig~2a is consistent with Fig~3c in~\cite{jotunable}, where an analytical approach is undertaken and the proportion of the FP is depicted as a function of degree for  synthetic networks.

Now let us focus on the solid black curve in Fig~2a which pertains to mean follower superiority. This proportion also increases initially, followed by a plateau that lasts up to degrees around 100, and then decreases. The decline is  steeper than the case of mean followee superiority. 

For the next example, we consider Fig~2h. The solid red curve starts at 0.3. It increases up to 0.9, with a plateau that lasts up to an  FTR of around 0.04, then decreases monotonically. For example, a user with an FTR of 0.01 has a higher proportion of experiencing mean followee superiority in FTR (proportion is around 0.9) 
than a user with an FTR of 0.001 (for whom the proportion is 0.7). 

The presence of positive slopes and/or plateaus is visible in most of the curves presented in Fig~2. Such a behavior is in stark contrast with one might intuitively expect to observe (a monotonically decreasing curve, as mentioned above). In Figs~2a,~2b,~2g, and~2h, the initial increase in the proportion of experiencing neighbor superiority indicates that those with minimum (and close to minimum) values of attributes establish links both amongst themselves and those with higher values of attributes. However, those with intermediate levels of the attributes  tend to establish links only towards those who have higher attributes than them. In the hierarchical representation of the system, we can say that nodes with close-to-minimum attributes connect both up and across, and nodes with higher levels of attributes only connect up. We will get back to this point below when we discuss Fig~3.

In Figs~2c,2d,~2e, and~2f, the curves do not exhibit a steep positive slope. Rather, they begin with plateaus, followed by steep decrease. This pattern suggests that   nodes with close-to-minimum levels of these attributes follow those with higher attributes than them. The hierarchies that pertain to these attributes are more upwardly-oriented; most nodes tend to connect up, rather than across.

In all the figures, for the same type of superiority, the median curve falls  below the mean curve. This implies that for any type of superiority, the proportion of experiencing median superiority is smaller than the proportion of experiencing mean superiority.

In Fig~2, different  numbers of nodes fall  into  different bins. This is true for all nodal attributes. This is caused by the high skew in the distributions of the nodal attributes. This results in loss of valuable information. For example, we know from Fig~2e that a user whose TTR equals 10 experiences mean followee superiority with proportion of almost 0.9, that is, almost  90\% of the users whose TTR is 10 experience mean followee superiority. However, this figure does not tell us where such users fall in the ranking of the TTR values.

 In Fig~3, we plot the proportion of experiencing different types of neighbor superiority as a function of the  nodes' percentile rank for different attributes. The horizontal axes represent the percentile ranks. For example, in the case of NOT which is depicted in Fig~3d, a percentile rank of 0.6 for a node means that 60\% of the nodes have NOT values smaller than or equal to the NOT of that node. In other words, the horizontal axes of Fig~3 results from a nonlinear rescaling of the horizontal axes of Fig~2. We have divided the interval between the minimum value and the maximum values for percentile ranks into 500 bins, and for nodes falling into the same bin, we calculated the fraction who experience the   types of neighbor superiority corresponding to that attribute. The curves in Fig~3 are more telling: we readily observe that in all the figures, there is a very wide plateau that stretches up to the very close proximity of percentile rank of 1. This strengthens the assertions made above about the hierarchical nature of connections: most users---even those with very high ranking of any attribute---are connected to those with higher attributes than them. Note that in some figures, the last bin does not have zero proportion of experiencing neighbor superiority. This is because the last bin stores the top 0.2\% of the population, and the proportion associated with this bin is averaged among the top 0.2\%. Since many of these users experience neighbor superiority, despite their very high ranking in the corresponding attributes, the average is not zero. 

Fig~4 reaffirms the hierarchical nature of the connections. The figure is depicted as follows. We first divide the range of in-degrees into 25 logarithmic bins and group the population accordingly. We then construct matrix A whose (i,j) element denotes the number of links that are from a node in bin j to a node in bin i. We then normalize the matrix A column-wise, so that each column sums up to unity. Let us denote the resulting matrix by B. Matrix B is depicted in Fig~4.  Column c represents the distribution of the destination of links whose starting points are nodes in bin c. The values on the bottom and left axes are the starting points of the bins. The values on the top and right axes are the corresponding percentile ranks of the starting points of the bins.

The matrix can be divided into four regions. For the nodes with in-degrees in the first five bins, the majority of outgoing links land on the nodes with highest in-degrees. For the nodes in the next nine bins the majority of the outgoing links point towards the nodes within the same bins. However, the fraction of links pointing to the nodes with larger in-degrees is higher than those pointing to nodes with smaller or equal in-degrees. The nodes in these two regions are likely to experience both the mean and median followee superiority. The nodes in the third region are mostly following the nodes with relatively high in-degrees (in the 98th and 99th percentiles), but not the highest bins. Finally, the nodes in the last four bins tend to follow nodes in the first seven bins. 

Note that nodes in the second region experience the followee superiority because of following nodes within the same region but with higher in-degrees; not for following the hubs. Moreover, if the structure of the graph was star-like, we would expect to see two dense regions in the top left and bottom right of the matrix. This would disregard the role of the nodes in the middle bins.


\begin{figure}[!h]
        \centering
        \begin{subfigure}[b]{0.5 \columnwidth}
                \includegraphics[width=\columnwidth, height=50mm]{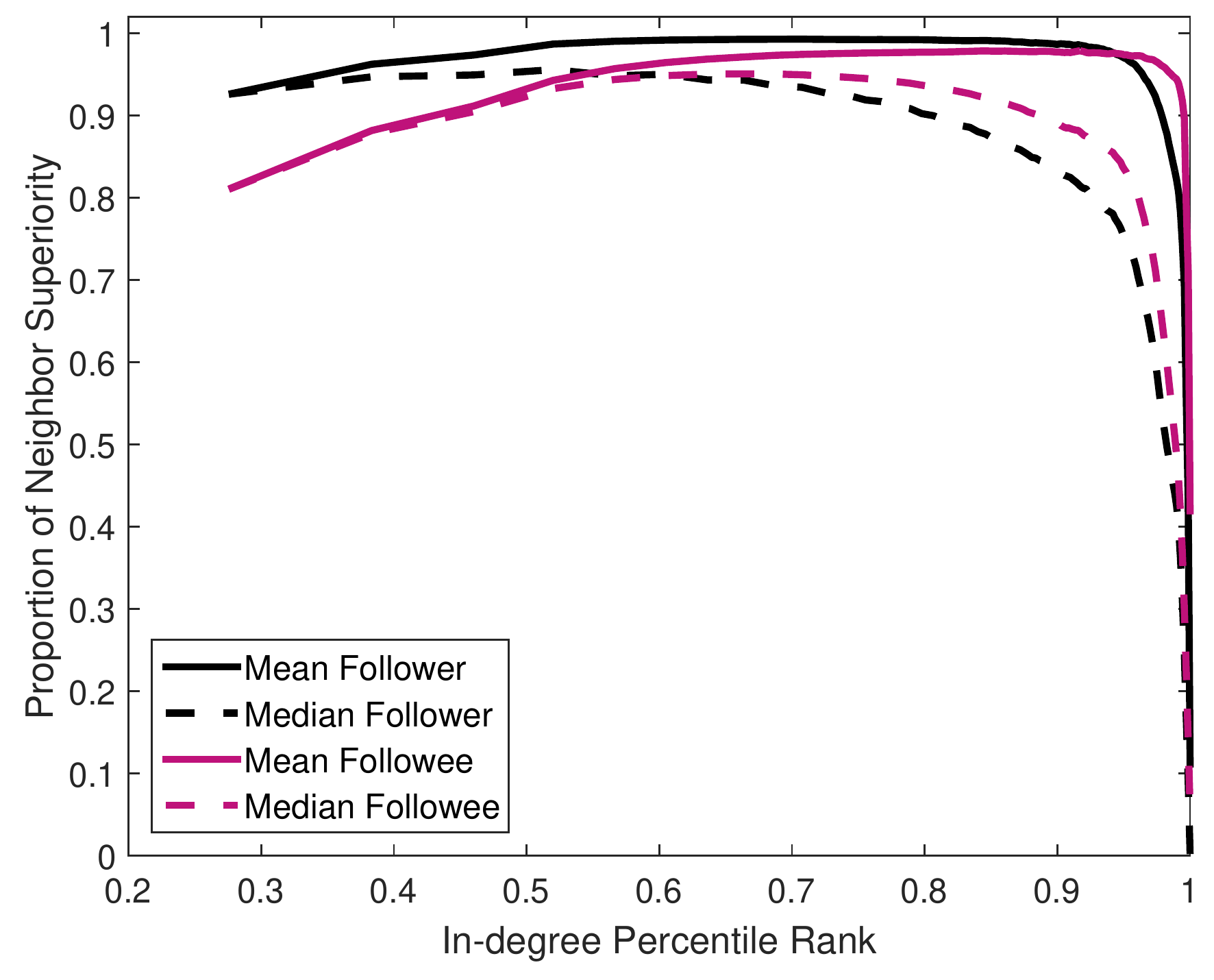}
                \caption{In-degree}
                \label{InDegreeP}
        \end{subfigure}%
        ~ 
        \begin{subfigure}[b]{0.5 \columnwidth}
                \includegraphics[width=\columnwidth, height=50mm]{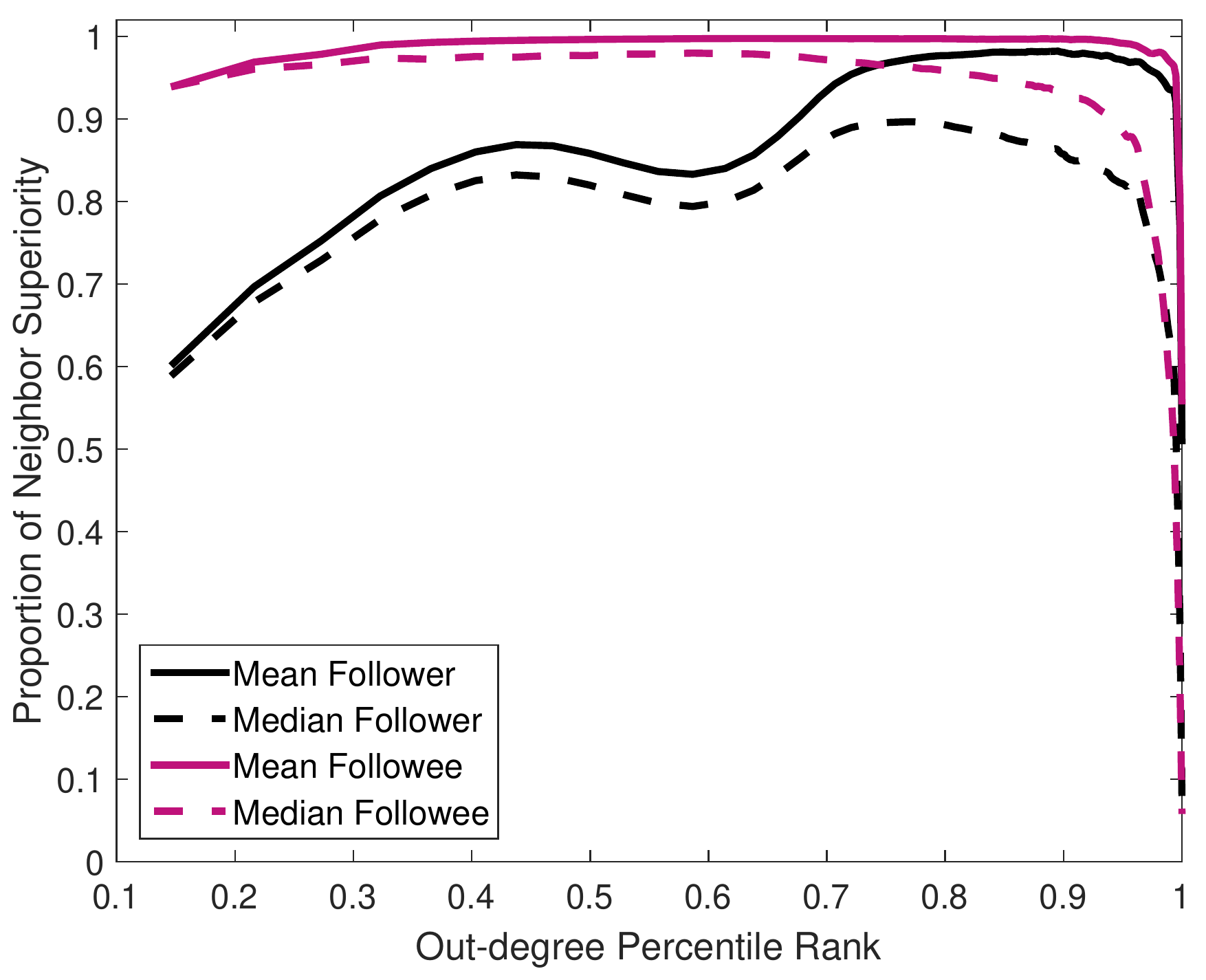}
                \caption{Out-degree}
                \label{OutdegreeP}
        \end{subfigure}
        \\
              \centering
        \begin{subfigure}[b]{0.5\textwidth}
                \includegraphics[width=\textwidth, height=50mm]{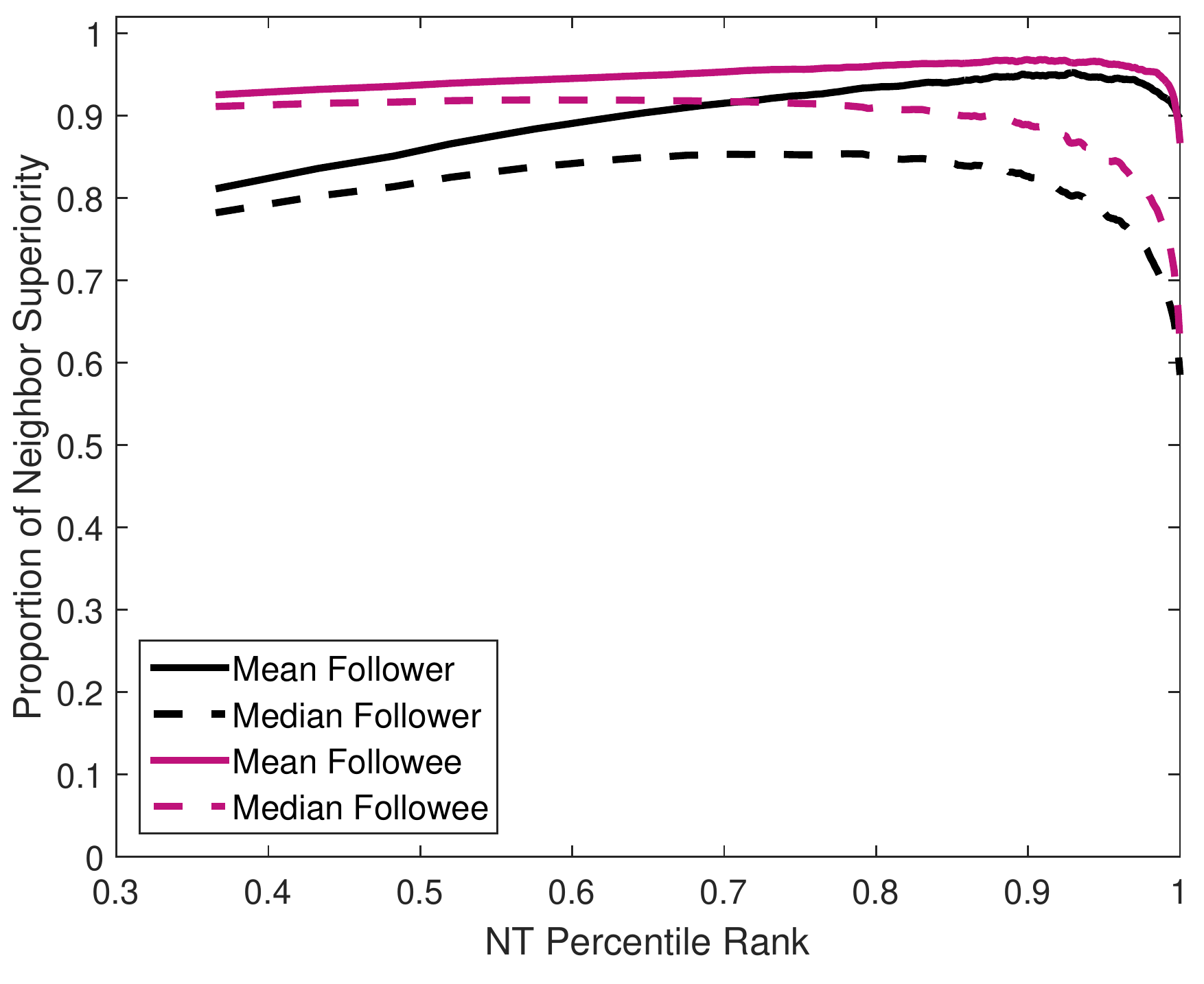}
                \caption{NT: Number of Tweets}
                \label{NTP}
        \end{subfigure}%
        ~ 
        \begin{subfigure}[b]{0.5\textwidth}
                \includegraphics[width=\textwidth, height=50mm]{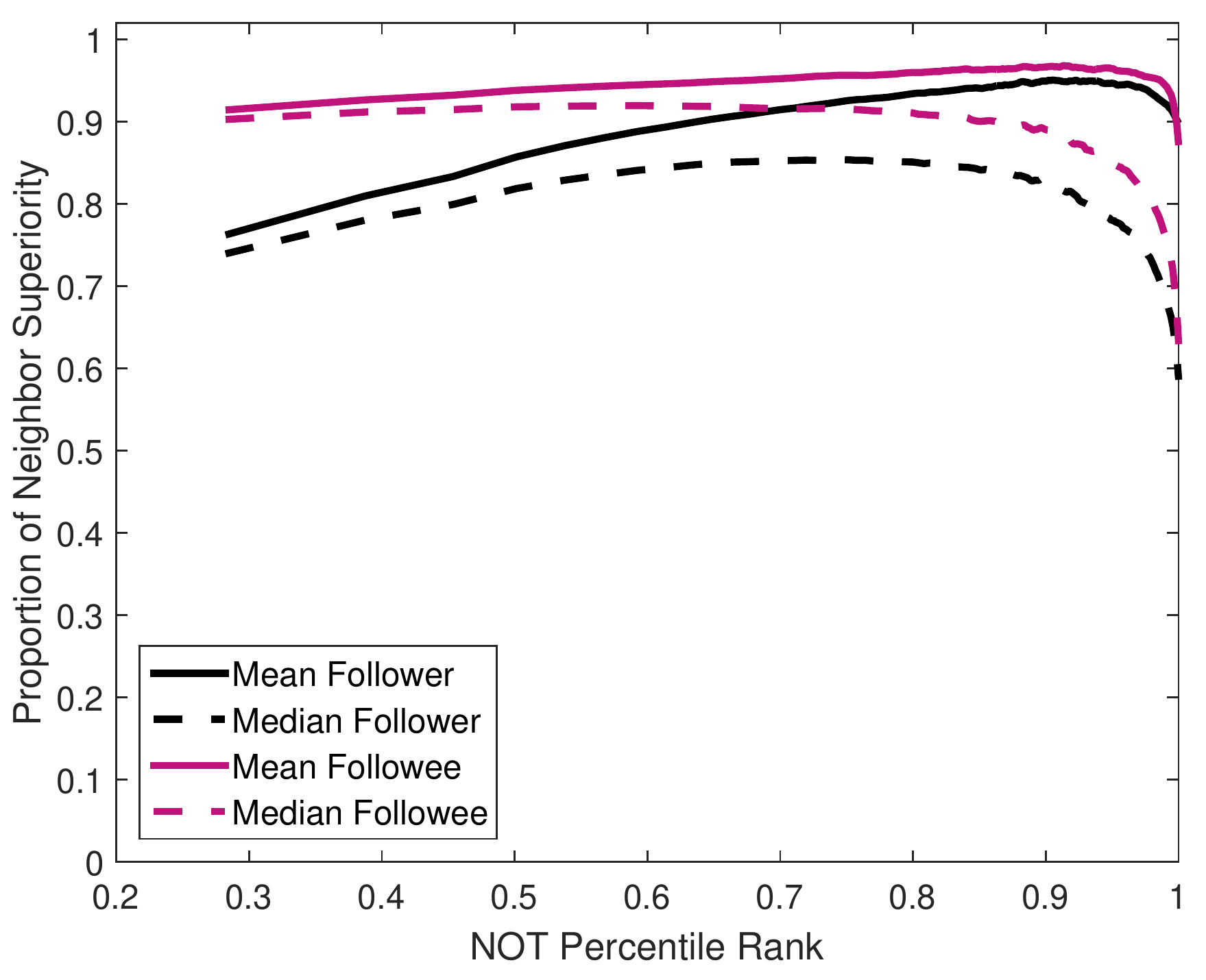}
                \caption{NOT: Number of Original Tweets.}
                \label{NOTP}
        \end{subfigure}
        \\
         \centering
        \begin{subfigure}[b]{0.5\textwidth}
                \includegraphics[width=\textwidth]{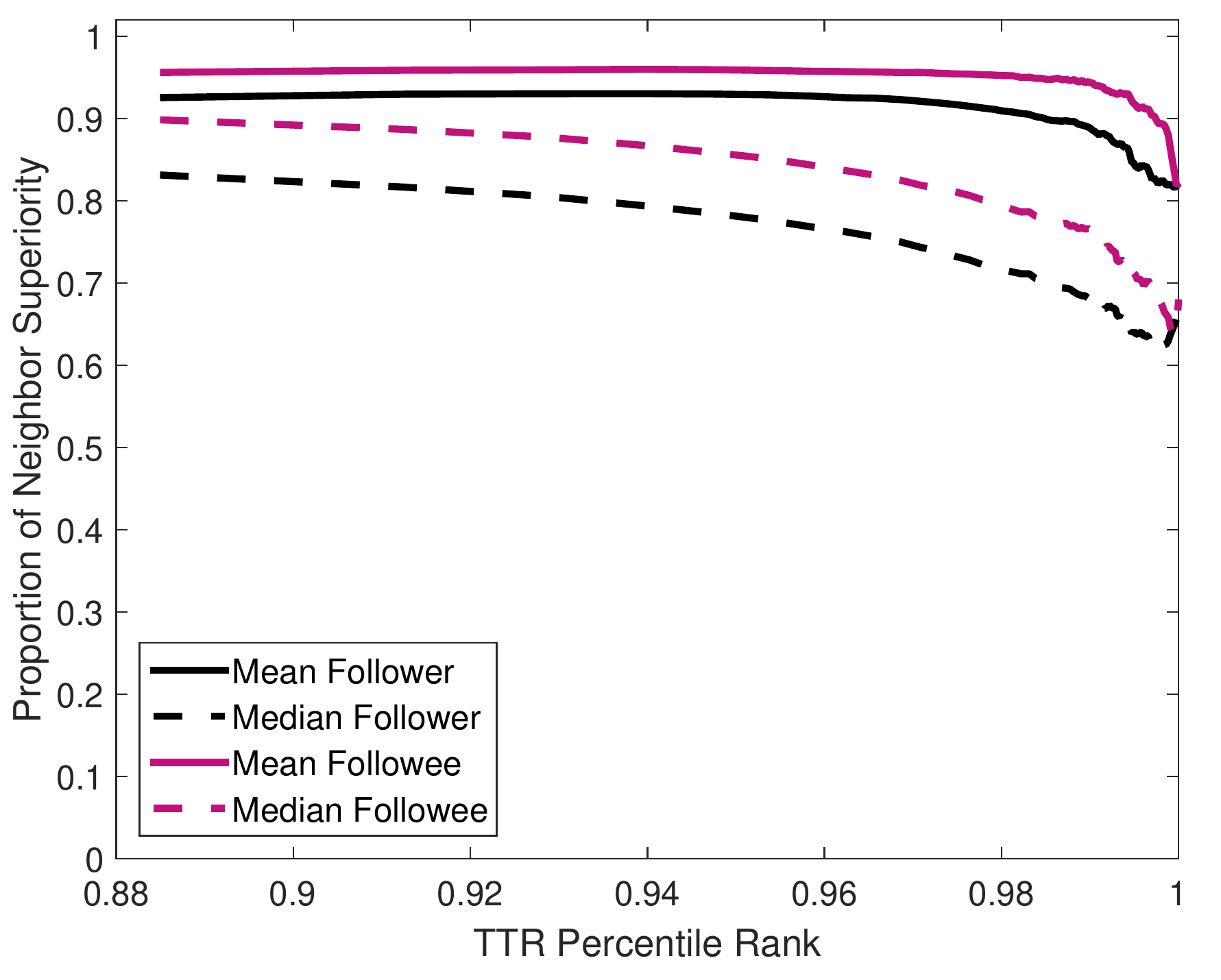}
                \caption{TTR: Total Times Retweeted.}
                \label{TTRP}
        \end{subfigure}%
        ~ 
        \begin{subfigure}[b]{0.5\textwidth}
                \includegraphics[width=\textwidth]{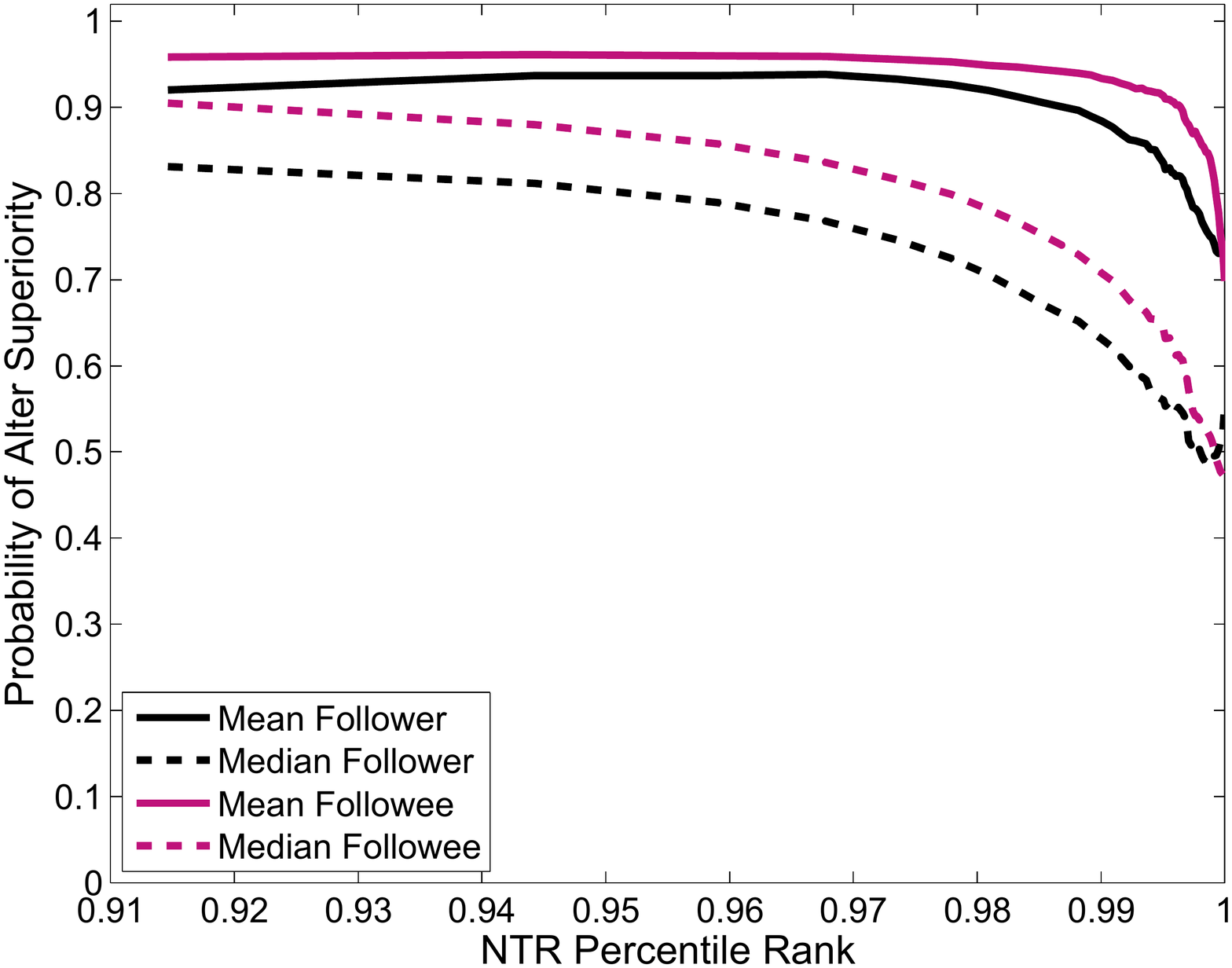}
                \caption{NTR: Number of Tweets Retweeted. }
                \label{NTRP}
        \end{subfigure}
        \\
         \centering
        \begin{subfigure}[b]{0.5\textwidth}
                \includegraphics[width=\textwidth]{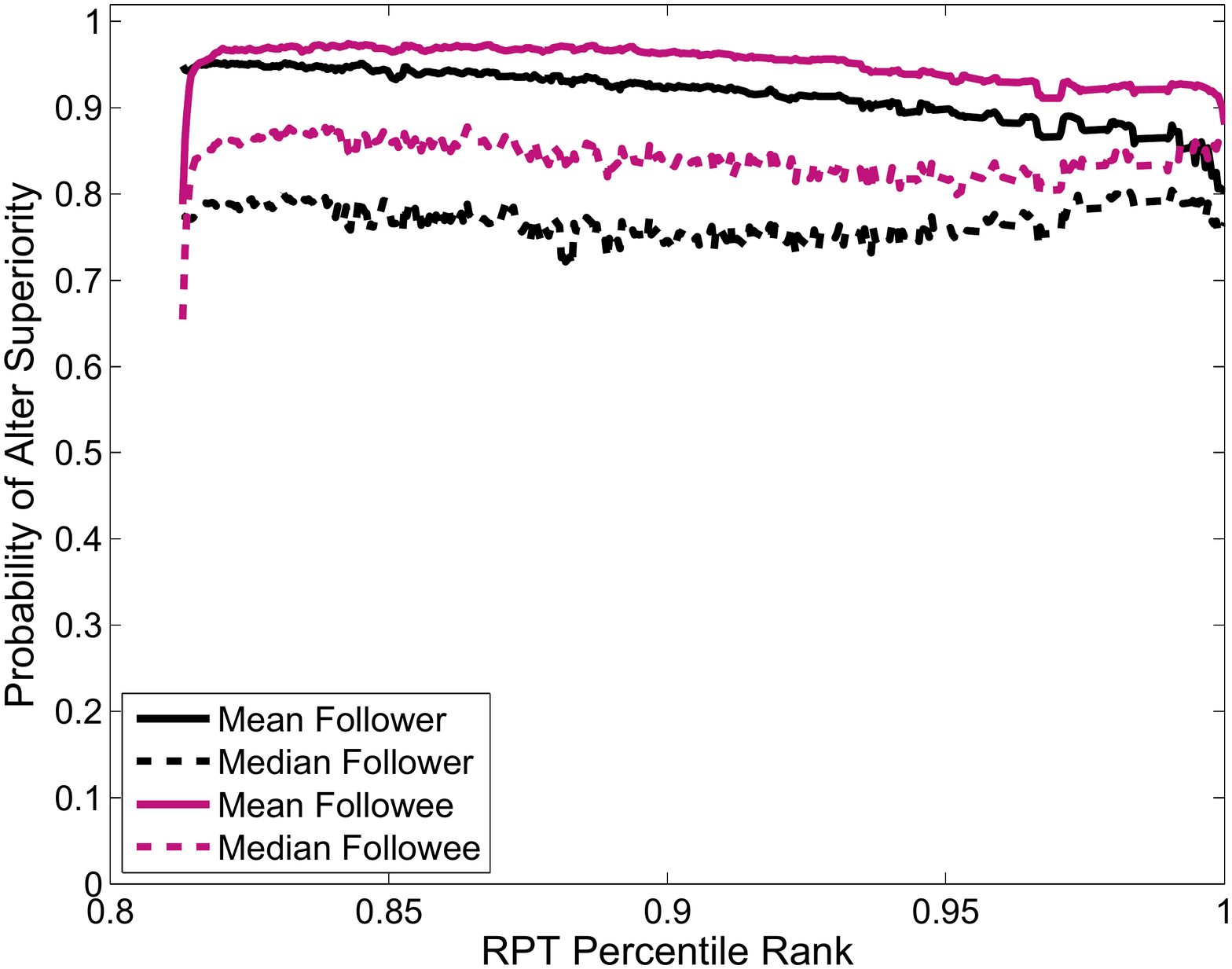}
                \caption{RPT: Retweets Per Tweet.}
                \label{RPTP}
        \end{subfigure}%
        ~ 
        \begin{subfigure}[b]{0.5\textwidth}
                \includegraphics[width=\textwidth]{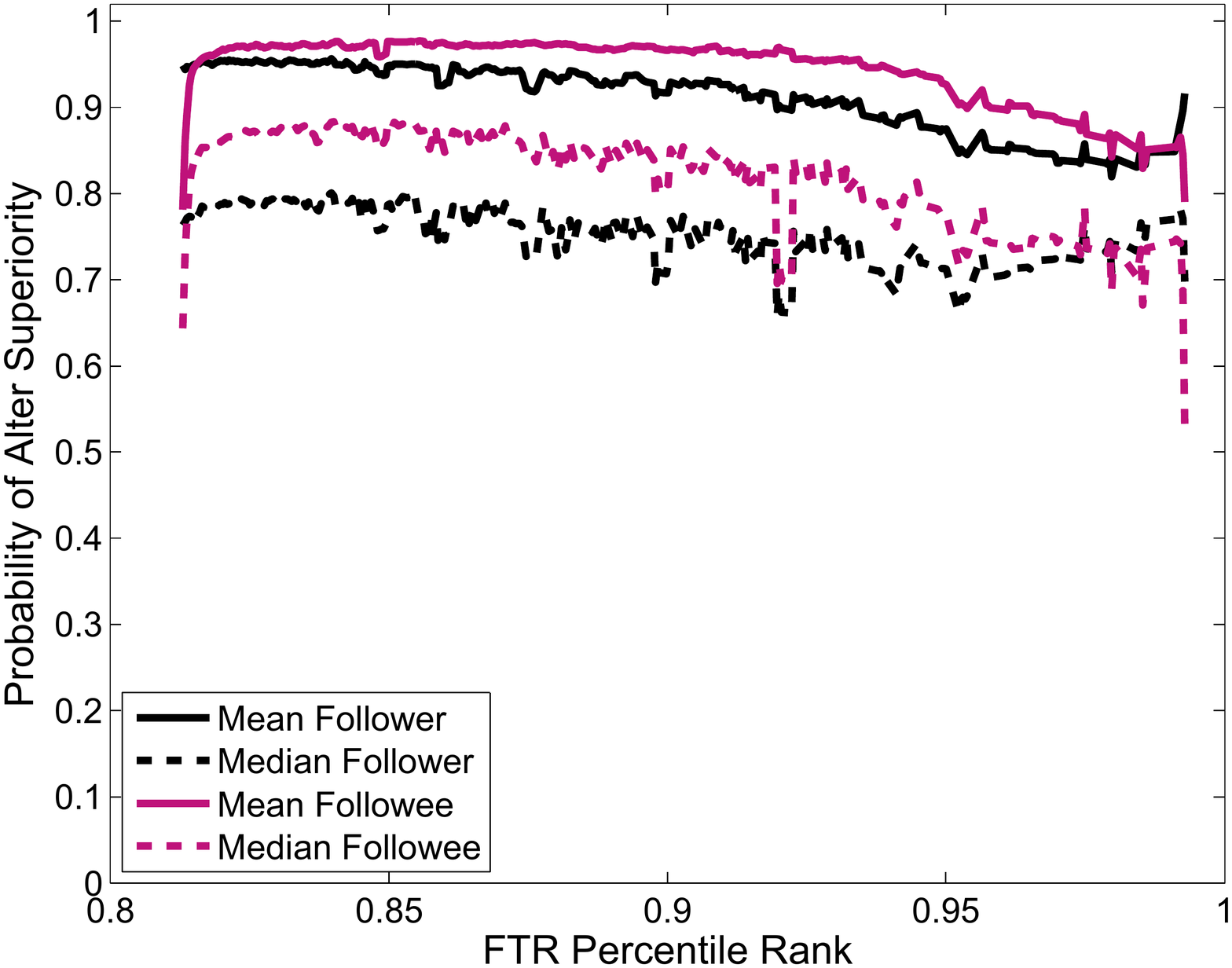}
                \caption{FTR: Fraction of Tweets Retweeted. }
                \label{FTRP}
        \end{subfigure}
        \caption{\footnotesize{Proportion of  neighbor superiority of different types for percentiles of nodal attributes.} }\label{fig:probP}
\end{figure}

 \begin{figure}[!h]
        \centering        
               \includegraphics[width=\columnwidth]{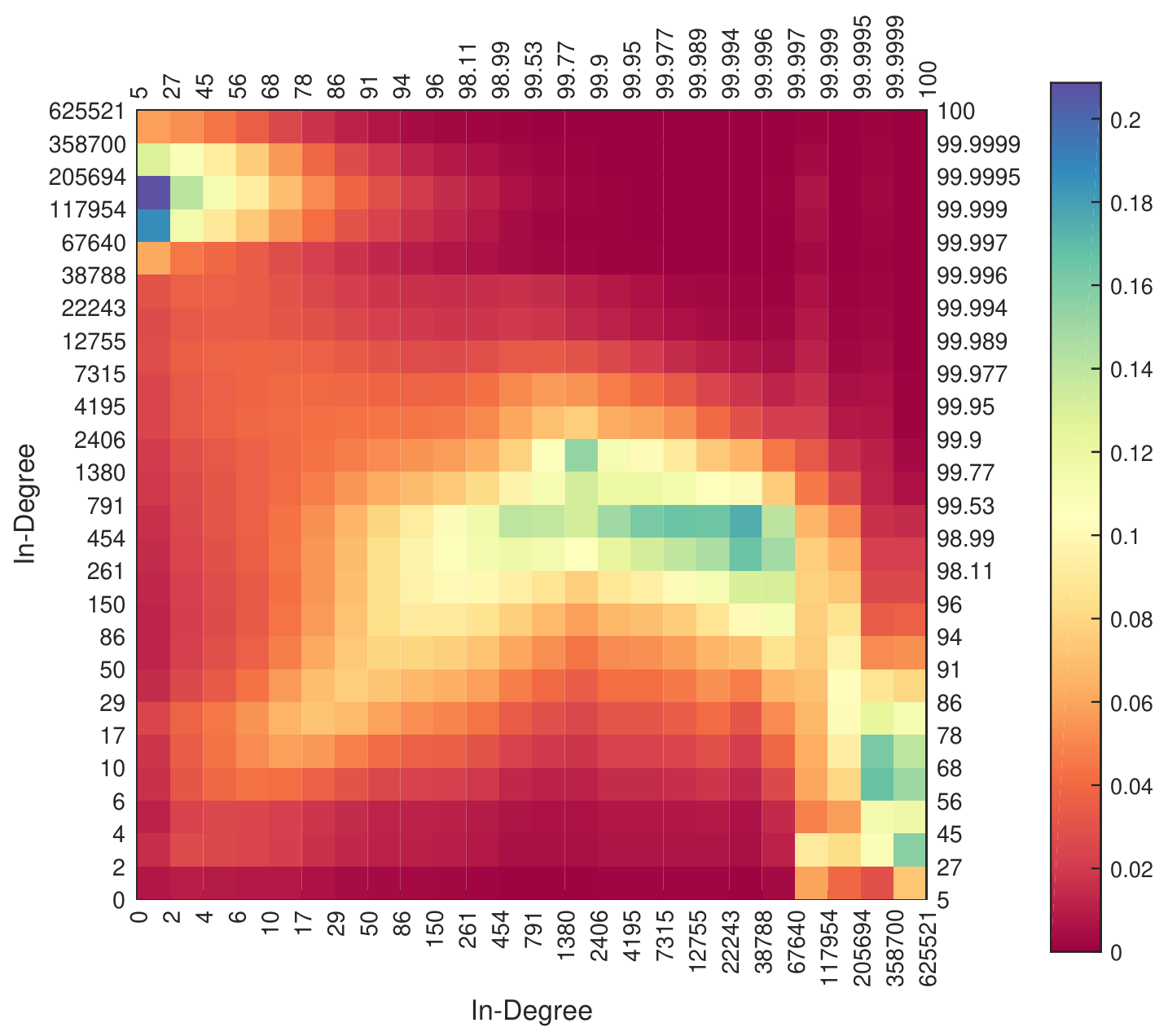}
                \caption{In-degree distribution of the followees of nodes as a function of their in-degree. The range of in-degrees are divided into 25 logarithmic bins. The values on the bottom and left axes are the starting points of the bins. The values on the top and right axes are the corresponding percentile ranks of the starting points of the bins.  Each column is normalized.}
                \label{graph}
 \end{figure}       

%
%

%
%

 \section*{Conclusion}

In this paper we introduced     six new measures to quantify different aspects of user activity and influence on social networks, and we computed them on a dataset of over 200 million tweets. We demonstrated
that the distributions of all of these attributes are heavy-tailed. Two of these attributes  (NT and NOT) are measures of activity, and four of them 
  (TTR, NTR, RPT and FTR) pertain to received retweets, and are measures of influence. The measures of influence are zero for more than 75\% of users, suggesting that the majority of Twitter users are observers of the content produced by a minority.
  
   We also introduced measures of neighbor superiority to quantify the local inequalities in different nodal attributes. We observed that the prevalence of mean neighbor superiorities of all types are above 63\%. 
    The prevalence of median neighbor superiorities of different types are also high; in 12 out of 16 types of median neighbor superiority, the prevalence is over 57\%.     
    We discussed that the  high prevalence of median versions of neighbor superiority  challenges the simplistic picture that neighbor superiority is a mere consequence of the existence of a few hubs in the network that put every peripheral node into experiencing neighbor superiority by elevating the average. 
   
   By inspecting  different types of neighbor superiority, we uncovered the hierarchical nature  of the connections in the Twitter graph both in terms of connectivity and in terms of nodal qualities. We observed that the fraction of nodes experiencing followee superiority exceeds the fraction of nodes experiencing follower superiority, and this is true for 15 out of 16 types of superiority  introduced.  This indicates the tendency of most users to follow other users who have higher attributes. It is of note that when  we speak of hierarchical structures, there are distinct hierarchies for different attributes. That is,  for example, if we once sort the node in terms of TTR, and then sort them in terms of in-degree, the hierarchies differ. Because the intra-node correlation between attributes are small (as presented in Table~\ref{correlations}) and therefore, a node that stands on the top of the hierarchy for TTR might be elsewhere  for in-degree. Let us point out that there are two distinct patterns of correlations: inter-node correlations for a given attribute, and intra-node correlation of different attributes. Our results indicate that hierarchies stem from high inter-node correlations of each given attribute.    
Also, our results indicate that intra-node correlation is not necessary for the individual-level GFP to hold in the case of Twitter. This agrees with the findings in~\cite{jotunable} for synthetic networks.

By close inspection of measures of neighbor superiority and the dependence of the likelihood of experiencing neighbor superiority on different attributes, we deduced that most users rarely follow down, rather, they tend to follow up or across, that is, they tend to follow other users with similar or higher attributes. This is true for almost every user, which makes even those in the top 0.5\% of the population experience neighbor superiority of different types. This means that Twitter does not possess a simple star-like structure, but is decentralized and inequalities exist locally for almost all nodes.

A counter-intuitive finding is that the proportion of experiencing neighbor superiority is not a monotonically-decreasing function of nodal attributes, or their ranks in those attributes. For example, it is not the case that the more re-tweets one receives, one's likelihood of experiencing neighbor superiority decreases. Rather, this likelihood is roughly constant up to the top percentile of the population in terms of retweets received. The trend is even reversed in the case of in-degree.  For example, the proportion of experiencing neighbor superiority can even increase as in-degree increases. To ensure low likelihood of experiencing neighbor superiority, it does not suffice to increase one's attribute; one needs to stand in a very high percentile of the population.

\nolinenumbers

%
%
%
\bibliography{refs}
%
%
%
%


\end{document}